\documentclass[twocolumn]{aastex63}
\usepackage{multirow}

\shorttitle{Estimate halo mass}
\shortauthors{Zhao et al.}
\graphicspath{{./}{figures/}}

\begin{document}

\title{From Halos to Galaxies. VI. Improved halo mass estimation for SDSS groups and measurement of the halo mass function }


\correspondingauthor{Yingjie Peng}
\email{yjpeng@pku.edu.cn}

\author[0009-0001-1564-3944]{Dingyi Zhao}
\affiliation{Department of Astronomy, School of Physics, Peking University, 5 Yiheyuan Road, Beijing 100871, People’s Republic of China}
\affiliation{Kavli Institute for Astronomy and Astrophysics, Peking University, 5 Yiheyuan Road, Beijing 100871, People’s Republic of China}

\author{Yingjie Peng}
\affiliation{Department of Astronomy, School of Physics, Peking University, 5 Yiheyuan Road, Beijing 100871, People’s Republic of China}
\affiliation{Kavli Institute for Astronomy and Astrophysics, Peking University, 5 Yiheyuan Road, Beijing 100871, People’s Republic of China}

\author[0000-0002-4534-3125]{Yipeng Jing}
\affiliation{Department of Astronomy, School of Physics and Astronomy, Shanghai Jiao Tong University, Shanghai 200240, People’s Republic of China}
\affiliation{Tsung-Dao Lee Institute, and Shanghai Key Laboratory for Particle Physics and Cosmology, Shanghai Jiao Tong University, Shanghai 200240, People’s Republic of China}

\author[0000-0003-3997-4606]{Xiaohu Yang}
\affiliation{Department of Astronomy, School of Physics and Astronomy, Shanghai Jiao Tong University, Shanghai 200240, People’s Republic of China}
\affiliation{Tsung-Dao Lee Institute, and Shanghai Key Laboratory for Particle Physics and Cosmology, Shanghai Jiao Tong University, Shanghai 200240, People’s Republic of China}

\author[0000-0001-6947-5846]{Luis C. Ho}
\affiliation{Kavli Institute for Astronomy and Astrophysics, Peking University, 5 Yiheyuan Road, Beijing 100871, People’s Republic of China}
\affiliation{Department of Astronomy, School of Physics, Peking University, 5 Yiheyuan Road, Beijing 100871, People’s Republic of China}

\author[0000-0002-7093-7355]{Alvio Renzini}
\affiliation{INAF--Osservatorio Astronomico di Padova, Vicolo dell’Osservatorio 5, I-35122 Padova, Italy}

\author[0000-0002-9656-1800]{Anna R. Gallazzi}
\affiliation{INAF--Osservatorio Astrofisico di Arcetri, Largo Enrico Fermi 5, I-50125 Firenze, Italy}

\author[0009-0000-7307-6362]{Cheqiu Lyu}
\affiliation{Department of Astronomy, School of Physics, Peking University, 5 Yiheyuan Road, Beijing 100871, People’s Republic of China}
\affiliation{Kavli Institute for Astronomy and Astrophysics, Peking University, 5 Yiheyuan Road, Beijing 100871, People’s Republic of China}
\affiliation{CAS Key Laboratory for Research in Galaxies and Cosmology, Department of Astronomy, University of Science and Technology of China, Hefei, Anhui 230026, China}
\affiliation{School of Astronomy and Space Science, University of Science and Technology of China, Hefei 230026, China}

\author[0000-0002-4985-3819]{Roberto Maiolino}
\affiliation{Cavendish Laboratory, University of Cambridge, 19 J.J. Thomson Avenue, Cambridge, CB3 0HE, UK}
\affiliation{Kavli Institute for Cosmology, University of Cambridge, Madingley Road, Cambridge, CB3 0HA, UK}
\affiliation{Department of Physics and Astronomy, University College London, Gower Street, London WC1E 6BT, UK}

\author[0000-0002-6961-6378]{Jing Dou}
\affiliation{School of Astronomy and Space Science, Nanjing University, Nanjing 210093, People’s Republic of China}

\author[0000-0002-0182-1973]{Zeyu Gao}
\affiliation{Department of Astronomy, School of Physics, Peking University, 5 Yiheyuan Road, Beijing 100871, People’s Republic of China}
\affiliation{Kavli Institute for Astronomy and Astrophysics, Peking University, 5 Yiheyuan Road, Beijing 100871, People’s Republic of China}

\author[0000-0002-3890-3729]{Qiusheng Gu}
\affiliation{School of Astronomy and Space Science, Nanjing University, Nanjing 210093, People’s Republic of China}

\author[0000-0002-4803-2381]{Filippo Mannucci}
\affiliation{INAF--Osservatorio Astrofisico di Arcetri, Largo Enrico Fermi 5, I-50125 Firenze, Italy}

\author[0000-0001-5356-2419]{Houjun Mo}
\affiliation{Department of Astronomy, University of Massachusetts, Amherst MA 01003, USA}

\author[0000-0002-6137-6007]{Bitao Wang}
\affiliation{Kavli Institute for Astronomy and Astrophysics, Peking University, 5 Yiheyuan Road, Beijing 100871, People’s Republic of China}

\author[0000-0003-1588-9394]{Enci Wang}
\affiliation{CAS Key Laboratory for Research in Galaxies and Cosmology, Department of Astronomy, University of Science and Technology of China, Hefei, Anhui 230026, China}
\affiliation{School of Astronomy and Space Science, University of Science and Technology of China, Hefei 230026, China}

\author[0000-0002-3775-0484]{Kai Wang}
\affiliation{Kavli Institute for Astronomy and Astrophysics, Peking University, 5 Yiheyuan Road, Beijing 100871, People’s Republic of China}

\author[0000-0002-8429-7088]{Yu-Chen Wang}
\affiliation{Department of Astronomy, School of Physics, Peking University, 5 Yiheyuan Road, Beijing 100871, People’s Republic of China}
\affiliation{Kavli Institute for Astronomy and Astrophysics, Peking University, 5 Yiheyuan Road, Beijing 100871, People’s Republic of China}

\author{Bingxiao Xu}
\affiliation{Kavli Institute for Astronomy and Astrophysics, Peking University, 5 Yiheyuan Road, Beijing 100871, People’s Republic of China}

\author[0000-0003-3564-6437]{Feng Yuan}
\affiliation{Center for Astronomy and Astrophysics and Department of Physics, Fudan University, Shanghai 200438, People’s Republic of China}

\author[0000-0002-9529-1044]{Xingye Zhu}
\affiliation{Department of Astronomy, School of Physics, Peking University, 5 Yiheyuan Road, Beijing 100871, People’s Republic of China}
\affiliation{Kavli Institute for Astronomy and Astrophysics, Peking University, 5 Yiheyuan Road, Beijing 100871, People’s Republic of China}



\begin{abstract}
In $\Lambda$CDM cosmology, galaxies form and evolve in their host dark matter (DM) halos. Halo mass is crucial for understanding the halo-galaxy connection. The abundance matching (AM) technique has been widely used to derive the halo masses of galaxy groups. However, quenching of the central galaxy can decouple the coevolution of its stellar mass and DM halo mass. Different halo assembly histories can also result in significantly different final stellar mass of the central galaxies. These processes can introduce substantial uncertainties in the halo masses derived from the AM method, particularly leading to a systematic bias between groups with star-forming centrals (blue groups) and passive centrals (red groups). To improve, we developed a new machine learning (ML) algorithm that accounts for these effects and is trained on simulations. Our results show that the ML method eliminates the systematic bias in the derived halo masses for blue and red groups and is, on average, $\sim1/3$ more accurate than the AM method. With careful calibration of observable quantities from simulations and observations from SDSS, we apply our ML model to the SDSS \citealp{2007ApJ...671..153Y} groups to derive their halo masses down to $10^{11.5}\mathrm{M_\odot}$ or even lower.  The derived SDSS group halo mass function agrees well with the theoretical predictions, and the derived stellar-to-halo mass relations for both red and blue groups match well with those obtained from direct weak lensing measurements. These new halo mass estimates enable more accurate investigation of the galaxy-halo connection and the role of the halos in galaxy evolution.
\end{abstract}

\keywords{galaxies: evolution ---formation ---dark matter halos}


\section{Introduction} \label{sec:intro}

Our understanding of galaxy formation and evolution is fundamentally based on the premise that all galaxies form and evolve within DM halos \citep{1978MNRAS.183..341W,1984Natur.311..517B}. The evolution of galaxies is significantly influenced by the hierarchical assembly of their host dark matter (DM) halos. Therefore, studying the relationships between various properties of galaxies and their host DM halos helps in understanding the evolution and formation of galaxies, constraining the cosmology parameters, and investigating the distribution and nature of DM  (see \cite{2018ARA&A..56..435W} for a comprehensive review on this topic). 

Mass is an important property of DM halos. There are primarily two categories of methods used to derive halo mass. The first category includes techniques designed for measuring the halo mass of individual galaxies or galaxy groups\footnote{All galaxies residing in the same DM halo constitute a galaxy group, even if it comprises only one galaxy.}. Examples of these are through measuring rotation curves in disk galaxies (e.g., \citealp{2001ARA&A..39..137S}), observing X-ray gas in giant elliptical galaxies (e.g., \citealp{2010ApJ...722...55N}), analyzing kinematics of globular clusters and planetary nebulae (e.g., \citealp{2003Sci...301.1696R,2009AJ....137.4956R}), estimating the globular cluster system mass (e.g., \citealp{2009MNRAS.392L...1S}), counting the number of the globular cluster systems (e.g., \citealp{2020AJ....159...56B}), fitting strong gravitational lensing (e.g., \citealp{2005ApJ...623L...5F}) and analyzing the kinematics of all galaxy members in DM halos (e.g., \citealp{2014A&A...566A...1T}). However, these methods have their limitations, such as the need for high spatial resolution to resolve individual globular clusters and the applicability of strong gravitational lensing only to massive galaxies and massive galaxy clusters. Consequently, only a limited number of individual galaxies or galaxy groups can be accurately measured for their host DM halo mass. The second category comprises statistical techniques such as weak gravitational lensing (e.g., \citealp{2005ApJ...635...73H,2006MNRAS.368..715M,2018ApJ...862....4L}) and the AM method (e.g., \citealp{2004ApJ...614..533T,2007ApJ...671..153Y}). Weak gravitational lensing statistically combines the lensing signals from numerous halos to deduce the characteristic halo mass for galaxy groups with specific properties. The AM method assumes that the most massive or luminous galaxy group occupies the most massive DM halo, followed by the second most massive or luminous galaxy group living in the second most massive halo, and so forth. Consequently, using a given halo mass function (HMF), halo mass can be assigned to galaxy groups based on their stellar mass or luminosity rank. A fundamental drawback of these approaches is their inability to ascertain precise halo masses for individual galaxy groups. The weak gravitational lensing method can only yield the average halo mass of a large sample of galaxies, and \cite{2016ApJ...819..119L} suggest that the scatter of halo mass estimated by the AM method in \cite{2007ApJ...671..153Y} may correlate with specific galaxy properties such as the specific star formation rate (sSFR) and star formation history (SFH).

Moreover, various studies confirm the biases observed in the AM method. Notably, investigations using the kinematics of satellite galaxies \citep{2011MNRAS.410..210M}, weak gravitational lensing \citep{2021A&A...653A..82B,2022A&A...663A..85Z,2024ApJ...960...71Z}, kinematics of the globular cluster systems and H\textsc{i} rotation curves \citep{2021A&A...649A.119P}, semi-analytical simulations \citep{2019ApJ...881...74M}, and hydrodynamical simulations \citep{2016MNRAS.463L...1C} all report that the red or passive galaxies reside in DM halos more massive than those hosting blue or star-forming galaxies with the same stellar mass.

This observed trend aligns with a scenario discussed in \cite{2010ApJ...721..193P} and \cite{2019ApJ...881...74M}, wherein the quenching of central galaxies results in different stellar-to-halo mass relations (SHMRs) between blue and red groups. In this scenario, blue groups experience growth in the stellar mass of their central galaxies primarily through star formation and mergers. Concurrently, their host DM halos also increase in mass through accretion and mergers, leading to a strong positive correlation between the stellar mass of their central galaxies and the mass of their host DM halos. Conversely, the stellar mass of central galaxies for red groups typically remains stable, growing only through occasional mergers, while their DM halos continue to grow. 
Consequently, at equivalent stellar mass of central galaxies, the DM halos of red groups are generally more massive than those of blue groups.

\cite{2009MNRAS.398L..58R} points out that two star-forming galaxies, which have the same stellar mass and slightly different sSFR at $z=2$, can exhibit a large difference in their stellar mass by $z=0$. If their DM halos assemble in a similar manner, their halo mass would be comparable at $z=0$. This challenges the AM method. 

The assembly history of DM halos also affects the SHMR. For instance, \cite{2017MNRAS.465.2381M} shows that in the EAGLE simulation, halos that formed later are more massive than those formed earlier at the same stellar mass, thereby introducing a bias in halo mass estimations via the AM method. This can also be found in L-GALAXIES \citep{2023ApJ...959....5L}. Additionally, \cite{2024arXiv240703409L} indicates that properties linked to the SFH of galaxies can predict the formation time of their DM halos. Thus, employing these SFH-related properties to estimate the masses of their host DM halos could lead to more accurate and unbiased measurements.

The primary objective of this paper is to develop a methodology that can be broadly applied across diverse galaxy populations to estimate their host DM halo mass with enhanced precision. \cite{2019ApJ...881...74M} applied a machine learning (ML) algorithm, the Random Forest regressor, to galaxy groups in the mock catalog based on the semi-analytical model L-GALAXIES to explore these nonlinear multicorrelations between halo mass and some easily acquired SFH-related galaxy properties and to verify the previous mentioned scenario. Their findings robustly support this scenario and indicate that halo mass can be more accurately inferred from these properties. Further, in this paper, we aim to minimize the discrepancies between the mock catalog and observational data, and subsequently train ML models on the mock catalog and apply these models to observational data to estimate halo mass. Fundamentally, the ML method aims to rectify the bias observed in the AM method by incorporating properties related to the SFH and/or merger history of galaxy groups. In this paper, the \emph{Planck} cosmology \citep{2014A&A...571A..16P} is adopted with the following cosmological parameters: $\sigma_8 = 0.829, H_0 = 67.3  \rm{kms}^{-1}\rm{Mpc}^{-1}, \Omega_\Lambda = 0.685, \Omega_m = 0.315, \Omega_b =0.04857$, and $n = 0.96$.

The structure of this paper is as follows. In Section \ref{sec:data}, we introduce the L-GALAXIES semi-analytical model and describe the mock catalog derived from it, which serves as the training, validation, and test sets. This section also details the group properties utilized in this study and discusses a group catalog based on SDSS DR7 used for predictions. In Section \ref{sec:method}, we explain our methods for calibrating the mock catalog with observational data, training the ML models on the mock catalog, and subsequently applying these models to estimate DM halo mass from observational data. In Section \ref{sec:test_set}, we evaluate the performance of the ML method within the test set. In Section \ref{sec:results}, we present the statistics of the halo masses derived from observational data using our ML method and compare these results with existing literature. We discussed the use of L-GALAXIES in Section \ref{sec:discussion}. Finally, we summarize our findings in Section \ref{sec:summary}.

\section{DATA} \label{sec:data}
\subsection{L-GALAXIES}\label{subsec:l-galaxies}

We use data from the publicly released Munich semi-analytical model, L-GALAXIES\footnote{\url{http://gavo.mpa-garching.mpg.de/MyMillennium}} \citep{2015MNRAS.451.2663H}. This model is based on the Millennium \citep{2005Natur.435..629S} and Millennium-II \citep{2009MNRAS.398.1150B} simulations. L-GALAXIES is rescaled to the \emph{Planck} cosmology \citep{2014A&A...571A..16P}.

Compared to hydrodynamic simulation, the semi-analytical model offers a computationally efficient approach to studying galaxy formation and evolution. It analytically describes the physical processes involved by utilizing DM halo merger trees. As a result, semi-analytical models typically cover a much larger volume, $(500\text{Mpc}/h)^3$ for L-GALAXIES, than hydrodynamic simulations. Therefore, we employ the mock catalog derived from the semi-analytical model to train our ML models, benefiting from its larger sample size, superior statistical robustness, and reduced cosmic variance.

L-GALAXIES has made several improvements relative to previous Munich galaxy formation models (e.g., \citealp{2011MNRAS.413..101G,2013MNRAS.428.1351G}). These improvements include delaying the return of material ejected by galactic winds, weakening ram-pressure stripping in low-mass halos, and lowering the gas surface density threshold for star formation. The model employs the Markov Chain Monte Carlo method to adjust the parameter space to match observations. Furthermore, L-GALAXIES yields one of the stellar mass functions that most closely matches observations. Additionally, it achieves reasonable matches across detailed distributions of color, sSFR, and luminosity-weighted stellar age for stars within the mass range of $8\leq\log(M_*/\mathrm{M_\odot})\leq 12$.

Given the aforementioned advantages, we employ the mock catalog generated by L-GALAXIES as the training, validation, and test sets for this study. Specifically, we select all galaxies with a stellar mass greater than $10^{9.0} \mathrm{M_\odot}$ from snapshots within the redshift range of $0.01 < z < 0.2$ in the mock catalog based on the Millennium simulation.

\subsection{Galaxy Group Properties}\label{subsec:group-property}
In this section, we detail the parameters selected for this study and explain the rationale behind these choices. As discussed in Section \ref{sec:intro}, the stellar-to-halo mass ratio is related to the SFH, merger history, and halo assembly history. Thus, the parameters chosen are closely associated with the SFH, merger history and/or halo assembly history. To ensure the applicability of our method to observational data, it is necessary that the parameters are available in both the L-GALAXIES model and observational datasets. Consequently, we have identified 25 parameters that satisfy these criteria, listed as follows: 

\begin{enumerate}
  \item Stellar mass of the central galaxy, $M_{\mathrm{cen}}$. The central galaxy is the most massive galaxies in its host DM halo.
  \item Total stellar mass of the galaxies more massive than the mass completeness threshold in the same group, $M_{\mathrm{tot}}$. The relationship between the total stellar mass and DM halo mass is tighter than that between the stellar mass of the central galaxy and DM halo mass (as demonstrated in  Figure 1  of \cite{2019ApJ...881...74M}). This is because $M_{\mathrm{tot}}$ contains more detailed information about the overall SFH and merging history of the group.
  \item The total number of galaxies more massive than the mass completeness threshold in the same group, richness \citep{2009ApJ...697.1842K}. Richness is directly linked to the merger history and serves as a reliable indicator of halo mass \citep{2012ApJ...746..178R}.
  \item Star formation rate (SFR) of the central galaxy. As previously discussed, the SHMR for blue and red groups varies significantly.
  \item $r$-band weighted stellar age of the central galaxy. We consider age as the weighted average of the time of star formation, i.e.,
  \begin{equation}
     age \sim \int_{0}^{t_0}  \mathrm{SFR}(t)(t_0-t)R(t)\,dt / L_r
  \end{equation}
  where $t_0$ is the age of the universe, $R(t)$ relates to the fraction of the stars formed at time $t$ that survive to the present and their $r$-band luminosity-to-mass ratio, and $L_r$ is the $r$-band luminosity. This parameter is tightly connected to the SFH.
  \item Color matrix of the central galaxies, $color$. This matrix includes all ten colors given by SDSS $u, g, r, i, z$ band absolute magnitudes. As discussed in Section \ref{sec:intro}, the SHMRs for central galaxies with various colors are different. 
  \item Total color matrix, $color_{\mathrm{tot}}$. These quantities are defined as the difference in total absolute magnitudes of group members more massive than the mass completeness threshold across $u, g, r, i, z$ bands. Analogous to the relationship between $M_{\mathrm{cen}}$ and $M_{\mathrm{tot}}$, $color_{\mathrm{tot}}$ contains more information than $color$.
\end{enumerate}

The mass completeness threshold mentioned above varies with redshift. At each specific redshift, only galaxies exceeding this threshold are selected, forming a volume-limited sample. The threshold is set higher at higher redshifts to accommodate observational constraints. The mass completeness threshold is detailed in Table \ref{tab:error} and indicated by the horizontal lines in Figure \ref{fig:samples}.

The virial mass of a DM halo is defined as the total mass enclosed within a sphere where the average density is 200 times the critical density of the universe.

We do not utilize the total SFR and total $r$-band weighted age due to the differing measurement error distributions between star-forming and passive galaxies. These differences complicate the estimation of errors when combining the data. The accuracy of measurement error distributions is crucial when applying the ML method to observational data, as discussed in Section \ref{sec:noise_importance}. Additionally, considering that using color as a feature for the ML model results in better performance than using magnitude, and that the information conveyed by absolute magnitudes can be effectively replaced by stellar mass and color, we exclude the SDSS $u, g, r, i, z$ band absolute magnitudes and total absolute magnitudes of group members more massive than the threshold from our analysis. 

\begin{deluxetable*}{clDcc}
\tablecaption{Parameters used to do data pre-processing \label{tab:error}}
\tablewidth{0pt}
\tablehead{
\colhead{Sample number} & \colhead{Redshift bin} & \multicolumn2c{Snapshot redshift} & \colhead{Median error of $M_{u},M_{g},M_{r},M_{i},M_{z}$ (mag)} & \colhead{log(Stellar mass threshold) ($\mathrm{M_\odot}$)} \\
}
\decimalcolnumbers
\startdata
0 & [0.01,0.04) & 0.025612 & 0.075, 0.013, 0.012, 0.015, 0.030 & 9.5 \\
1 & [0.04,0.07) & 0.053316 & 0.095, 0.013, 0.011, 0.013, 0.030 & 10.2 \\
2 & [0.07,0.1) & 0.082661 & 0.127, 0.015, 0.012, 0.014, 0.034 & 10.5 \\
3 & [0.1,0.13) & 0.11378 & 0.162, 0.017, 0.013, 0.015, 0.037 & 10.7 \\
4 & [0.13,0.165) & 0.147548 & 0.200, 0.020, 0.014, 0.015, 0.040 & 11 \\
5 & [0.165,0.2] & 0.183387 & 0.270, 0.025, 0.017, 0.017, 0.045 & 11.2 \\
\enddata

\tablecomments{We split the group sample into six subsamples, assigning each a sample number from zero to five. Each column details: (1) the sample number, (2) the corresponding redshift range in SDSS observation, (3) the corresponding snapshot redshift in the mock catalog, (4) the median measurement errors of the absolute magnitudes within each redshift range (also the standard deviation, $\sigma_0$, of the Gaussian noise added to the mock catalog), (5) the mass completeness threshold use to calculate $M_{\mathrm{tot}}$, richness, and $color_{\mathrm{tot}}$, which also defines whether galaxies belong to complete or incomplete samples.}
\end{deluxetable*}

\subsection{Galaxy Group Catalog}\label{subsec:group-catalog}

To apply the ML method to observational data, we require a group catalog to identify which galaxies belong to the same galaxy group, i.e., are embedded within the same DM halo. We utilize the most widely used group catalog in the local universe, the updated group catalog\footnote{\url{https://gax.sjtu.edu.cn/data/Group.html}} of \cite{2007ApJ...671..153Y}, which is constructed from NYU-VAGC DR7\footnote{\url{http://sdss.physics.nyu.edu/vagc/}} \citep{2005AJ....129.2562B} using a refined adaptive halo-based group finder \citep{2005MNRAS.356.1293Y}. Extensive testing in mock catalogs has demonstrated that this group finder outperforms traditional friends-of-friends algorithms. This catalog includes over 400,000 galaxy groups. We use their group catalog based on \texttt{model} magnitudes and SDSS redshift.

We extract stellar masses, SFRs, and their errors from the publicly available MPA-JHU DR7 of SDSS spectral measurements\footnote{\url{https://wwwmpa.mpa-garching.mpg.de/SDSS/DR7/}}. The errors are defined as the $(P84-P16)/2$ for their respective probability distribution functions (PDFs). Here, $P16$ and $P84$ represent the 16th and 84th percentiles of the PDFs, respectively. We use the median of these errors as the typical error. The typical error for stellar masses is 0.1 dex. For SFRs, the typical error is 0.3 dex for star-forming galaxies and 0.7 dex for passive galaxies. Stellar masses are determined by fitting photometry using the Bayesian methodology described by \cite{2003MNRAS.341...33K}. In-fiber SFRs, derived from H$\alpha$ emission \citep{2004MNRAS.351.1151B}, are aperture-corrected using photometry \citep{2007ApJS..173..267S} to estimate total SFRs. It is important to note the initial mass function (IMF) used in the MPA-JHU DR7 is the Kroupa IMF \citep{2001MNRAS.322..231K}, which differs from the Chabrier IMF \citep{2003PASP..115..763C} used in L-GALAXIES. Following the precedent set by \cite{2010MNRAS.408.2115M}, we amplify the stellar mass in L-GALAXIES by a factor of 1.06 to align with the observational data.

We derive the $r$-band weighted ages and their errors using the methodology outlined by \cite{2005MNRAS.362...41G}, which involves fitting five spectral absorption features simultaneously to estimate ages. The typical error of the ages is 1.1 Gyr for star-forming galaxies and 1.9 Gyr for passive galaxies.

We obtain the $u,g,r,i,z$ absolute magnitudes of galaxies and their errors from the NYU-VAGC DR7 \texttt{k-corrections} table\footnote{\url{http://sdss.physics.nyu.edu/vagc/kcorrect.html}}. These magnitudes are then rescaled to the cosmology parameters used in this study. The typical errors of absolute magnitudes across different redshift ranges are detailed in Table \ref{tab:error}.

Finally, we only select galaxies more massive than $10^{9.5} \mathrm{M_\odot}$. This is different from our previous selection of galaxies more massive than $10^{9.0} \mathrm{M_\odot}$ in the mock catalog, because we will add noise to the mock catalog and then select galaxies with masses greater than $10^{9.5} \mathrm{M_\odot}$, as shown in Section \ref{subsubsec:adding_noise}.


\section{METHOD} \label{sec:method}
In this section, we describe the calibration process for both the SDSS data and the mock catalog, aiming to harmonize their distributions. We also detail the machine learning process used in our analysis.

\subsection{Data Calibration}\label{subsec:data-prep}
If the distribution of group properties in the mock catalog is different from that in the SDSS data, applying a model trained on the mock catalog to the SDSS data will introduce unknown biases. Therefore, this calibration is crucial for accurately applying the model trained on the mock catalog to estimate DM halo mass in the SDSS data. To align the SDSS data and the mock catalog, we deal with the sample selection effect, add noise to the mock catalog, and normalize the galaxy properties of red and blue galaxy groups separately.

\begin{figure}[t]
  \centering
  \includegraphics[scale=0.45]{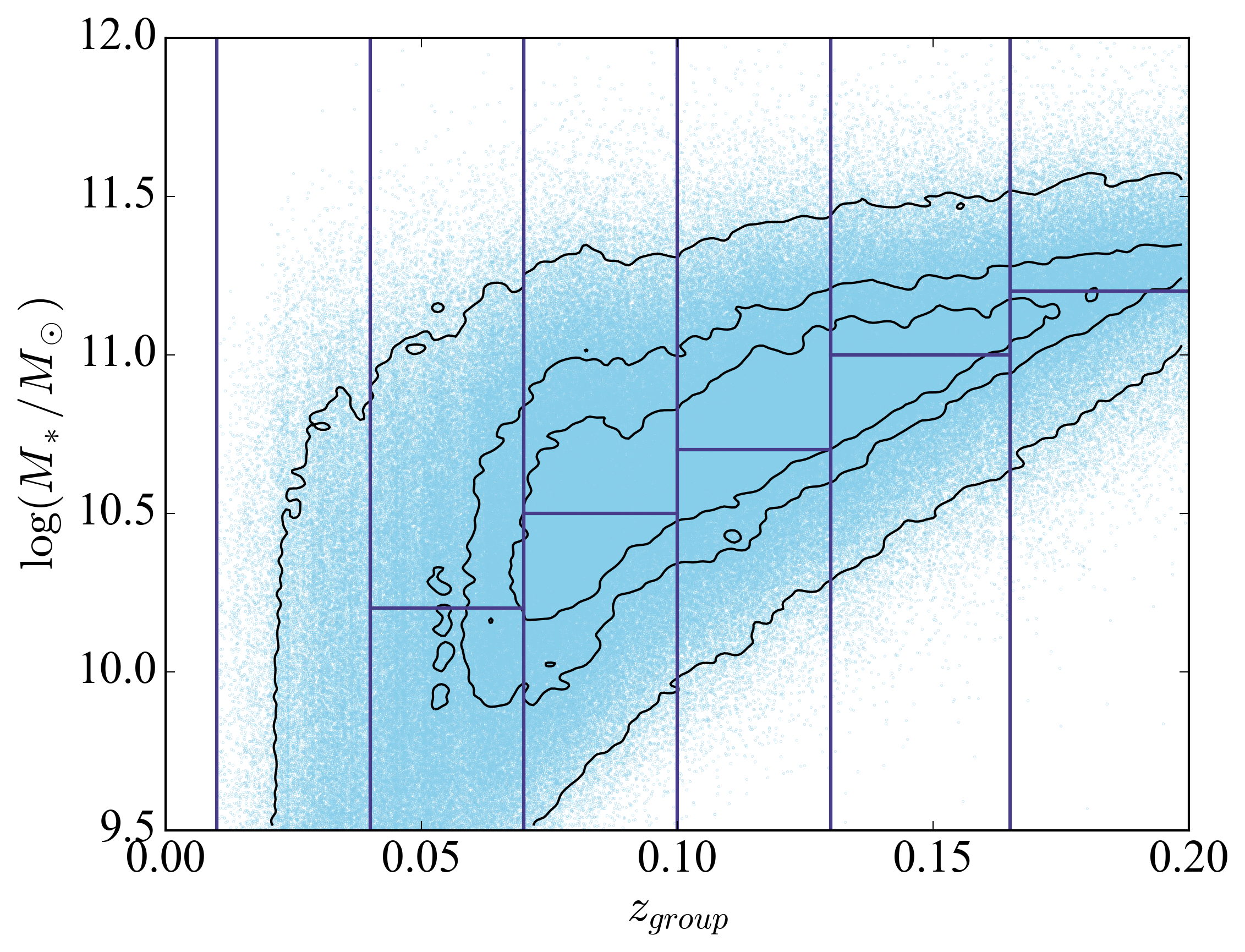}
  \caption{This figure shows the distribution of galaxies based on stellar mass and their host group redshift in the SDSS data. Blue dots represent individual galaxies, while black contours indicate the number density contours that encompass 35\%, 60\%, and 95\% of the galaxies. Five vertical lines divide the galaxies into six redshift ranges, corresponding to the six snapshots in the mock catalog, as detailed in Table \ref{tab:error}. Additionally, five horizontal lines divide the galaxies in the five redshift ranges above $z=0.04$ into two subsets, as specified in Table \ref{tab:error}. For each redshift bin above $z=0.04$, galaxies above the horizontal line form the complete sample, whereas those below the horizontal line constitute the incomplete sample. Galaxies with group redshifts below 0.04 belong to the complete sample.}
  \label{fig:samples}
\end{figure}
\begin{figure*}[t]
  \centering
  \includegraphics[scale=0.35]{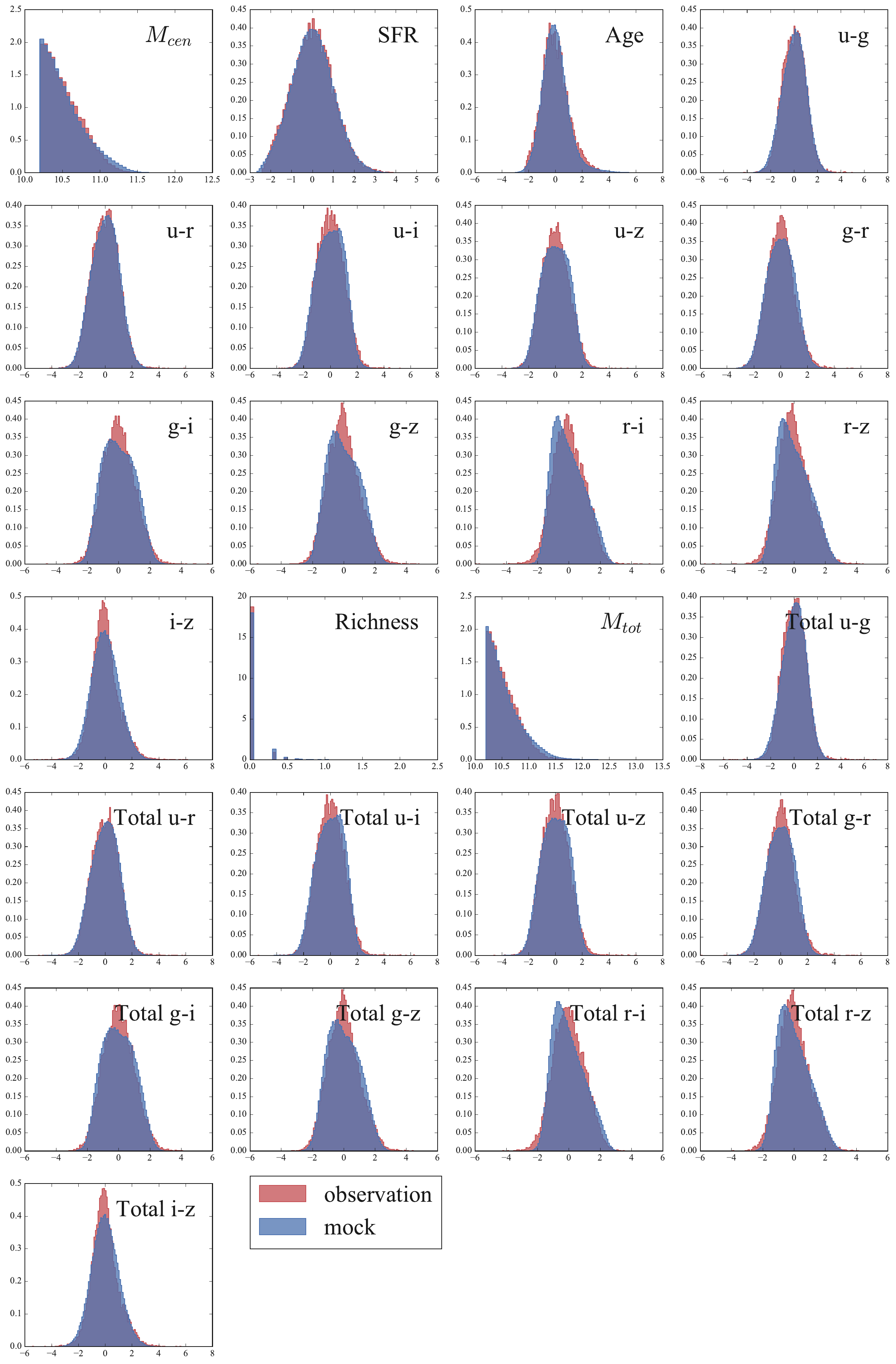}

  \caption{Distributions of the calibrated properties for blue groups in the complete sample within the redshift range $0.04<z<0.07$. Red shaded areas represent the distributions of group properties in the SDSS data, while blue shaded areas depict those in the mock catalog. Gaussian noise with a standard deviation of $\sigma_0$ has been added to the mock catalog. The richness is displayed on logarithmic scale. $M_{\mathrm{cen}}$ and $M_{\mathrm{tot}}$ are measured in $\mathrm{M_\odot}$. The units of the properties, which have been normalized, are arbitrary.}
  \label{fig:com_distribution_blue}
\end{figure*}
\begin{figure*}[t]
  \centering
  \includegraphics[scale=0.35]{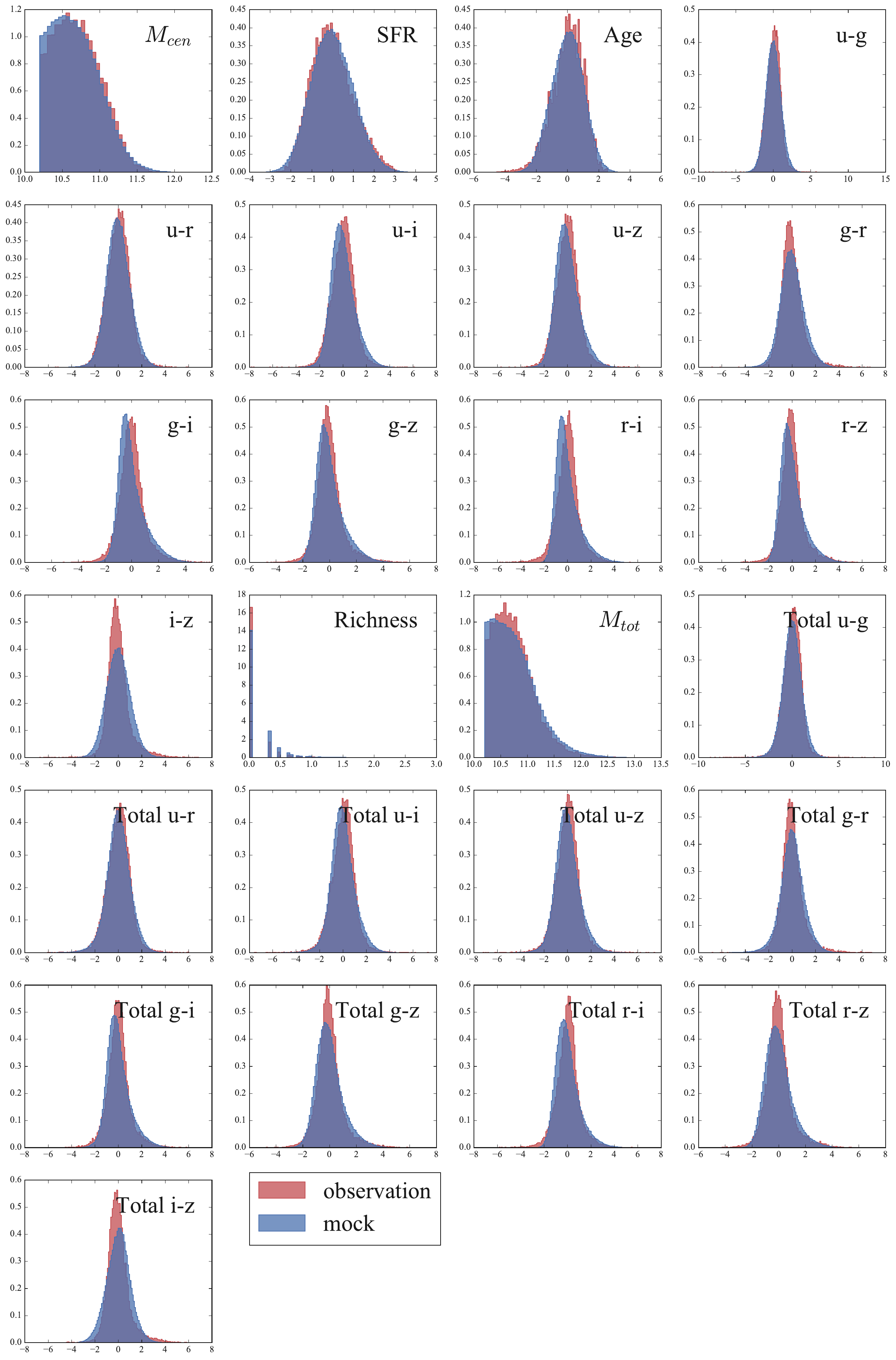}

  \caption{Distributions of the calibrated properties for red groups in the complete sample within the redshift range $0.04<z<0.07$. Similar to Figure \ref{fig:com_distribution_blue}, but focuses on the red groups in the complete sample.}
  \label{fig:com_distribution_red}
\end{figure*}
\begin{figure*}[t]
  \centering
  \gridline{\fig{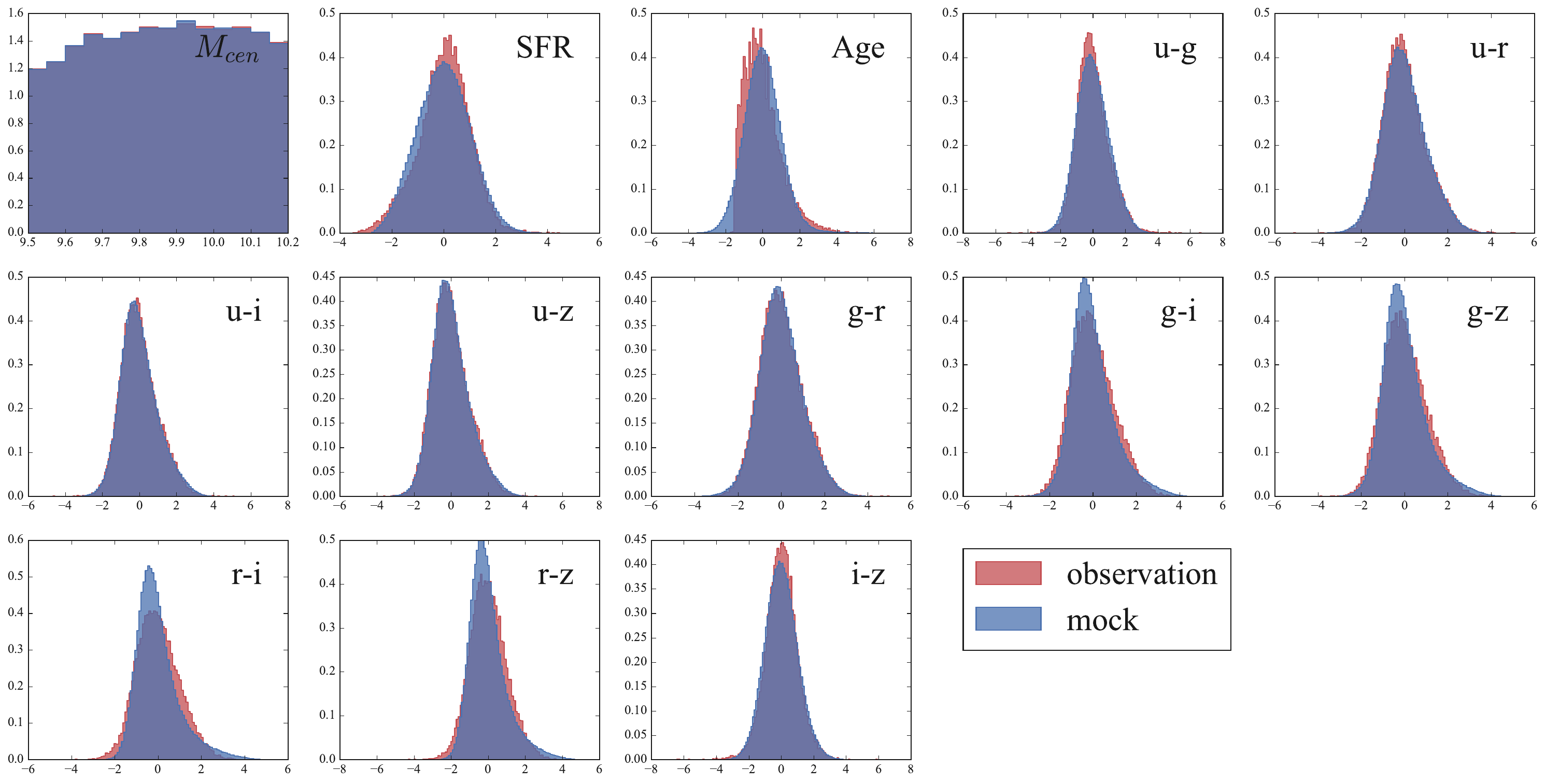}{0.95\textwidth}{Blue groups}}
  \gridline{\fig{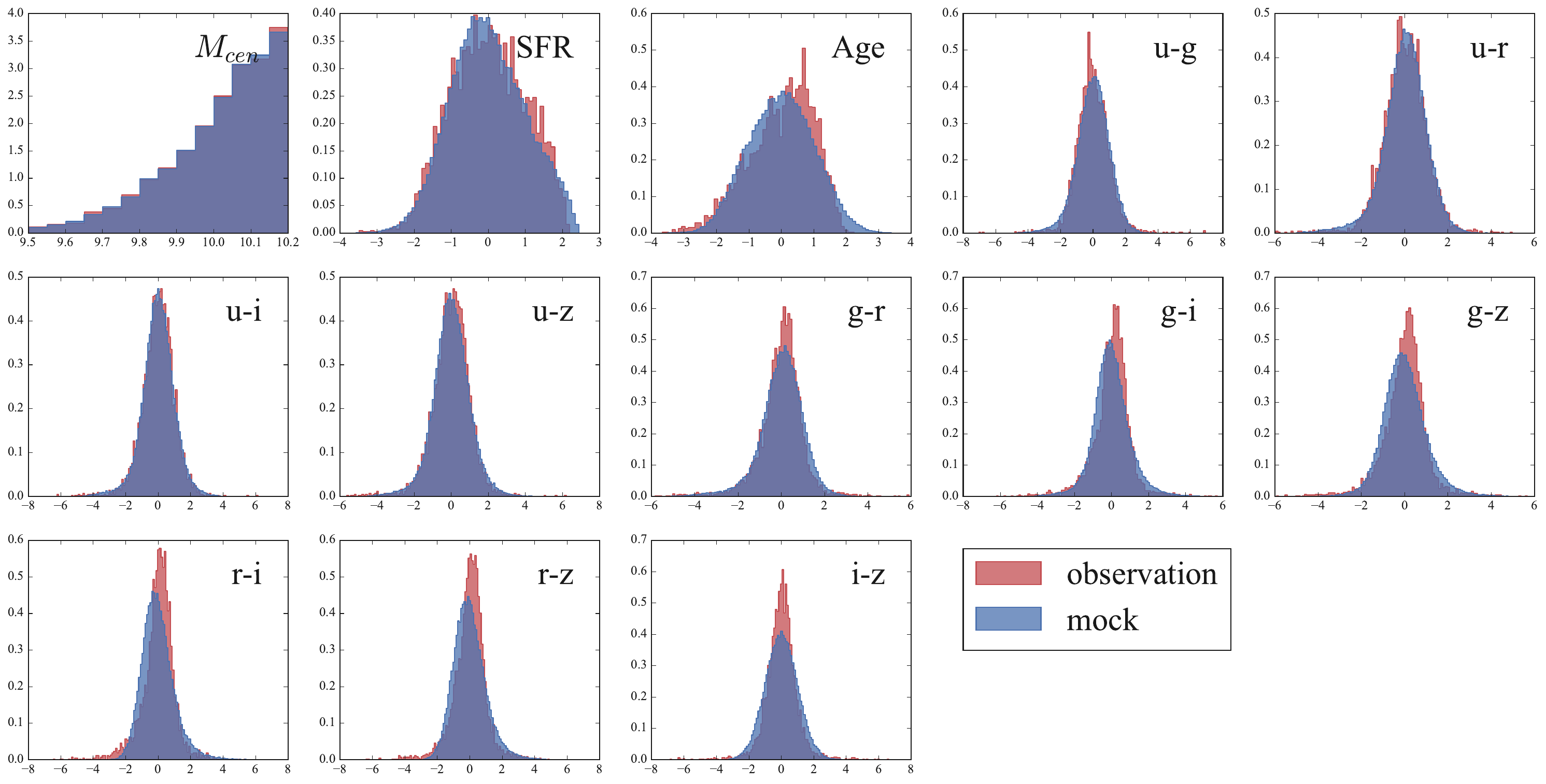}{0.95\textwidth}{Red groups}}
  \caption{Distributions of the calibrated properties for blue and red groups in the incomplete sample within the redshift range $0.04<z<0.07$. Similar to Figure \ref{fig:com_distribution_blue}, but focuses on the blue (upper panel) and red (lower panel) groups in the incomplete sample.}
  \label{fig:incom_distribution}
\end{figure*}
\begin{figure*}[htbp]
  \centering
  \includegraphics[scale=0.9]{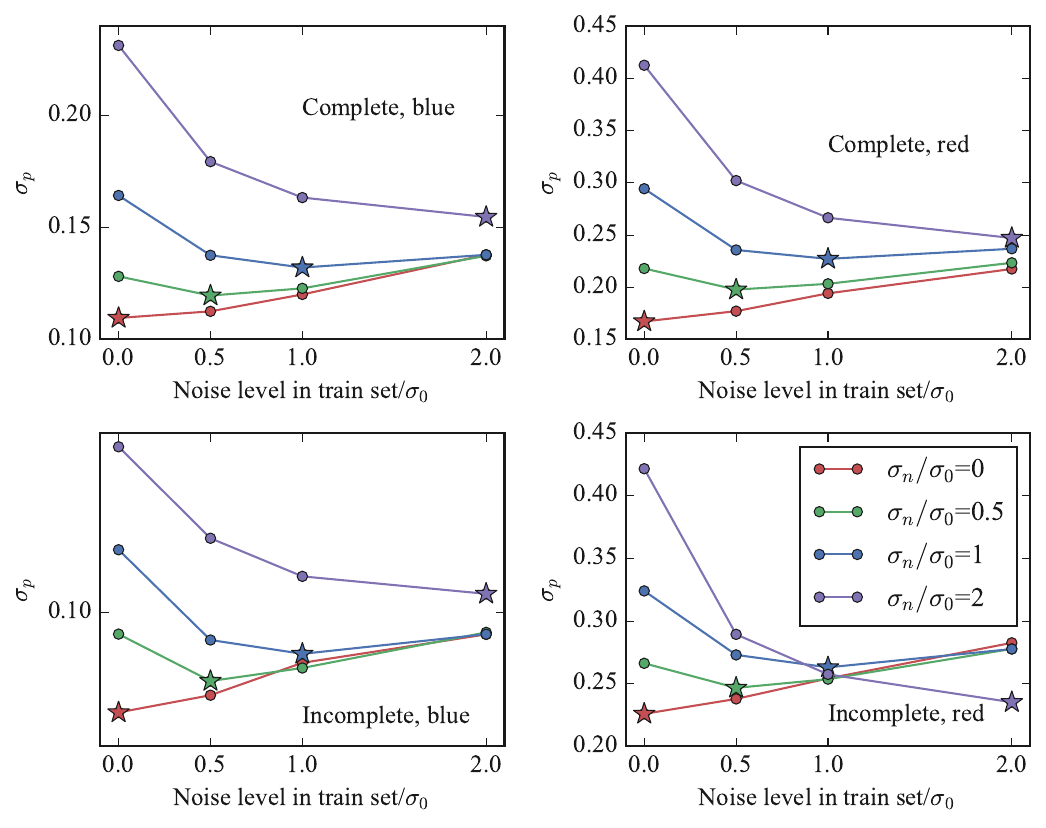}
  \caption{This figure illustrates the accuracy of ML models trained on training sets with varying levels of noise when applied to test sets with different noise levels. The x-axis represents the noise level in the training set, while the y-axis shows the standard deviation of the prediction residuals of halo masses. Different curves represent test sets with noise levels of $\sigma_n/\sigma_0 =$ 0 (red), 0.5 (green), 1 (blue), and 2 (purple). The stars highlight the points where the noise level in the training set matches that in the test set. \textbf{Top left panel}: Results for blue groups in the complete sample (sample number 1). \textbf{Top right panel}: Results for red groups in the complete sample. \textbf{Bottom left panel}: Results for blue groups in the incomplete sample. \textbf{Bottom right panel}: Results for red groups in the incomplete sample. All panels consistently show that the highest accuracy is achieved when the noise level in the training set matches that of the test set, as indicated by the stars.
  }
  \label{fig:noise_com}
\end{figure*}
\begin{figure*}[t]
  \gridline{\fig{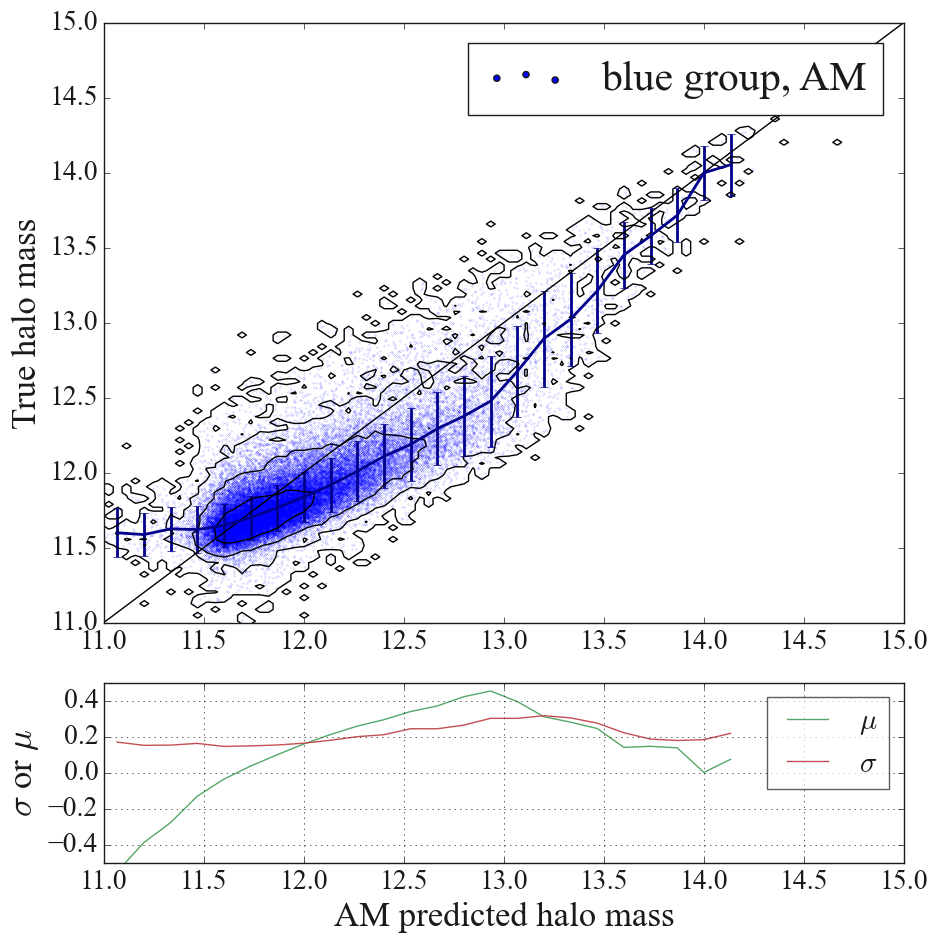}{0.5\textwidth}{}\fig{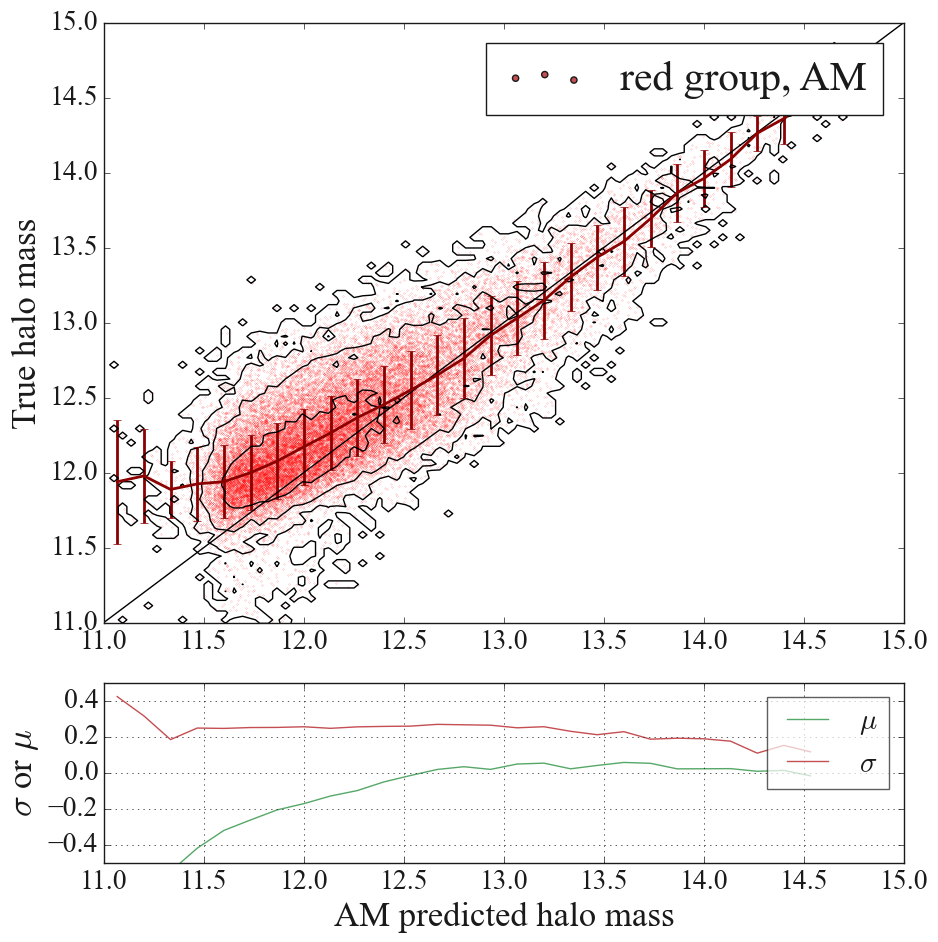}{0.5\textwidth}{}}
  \gridline{\fig{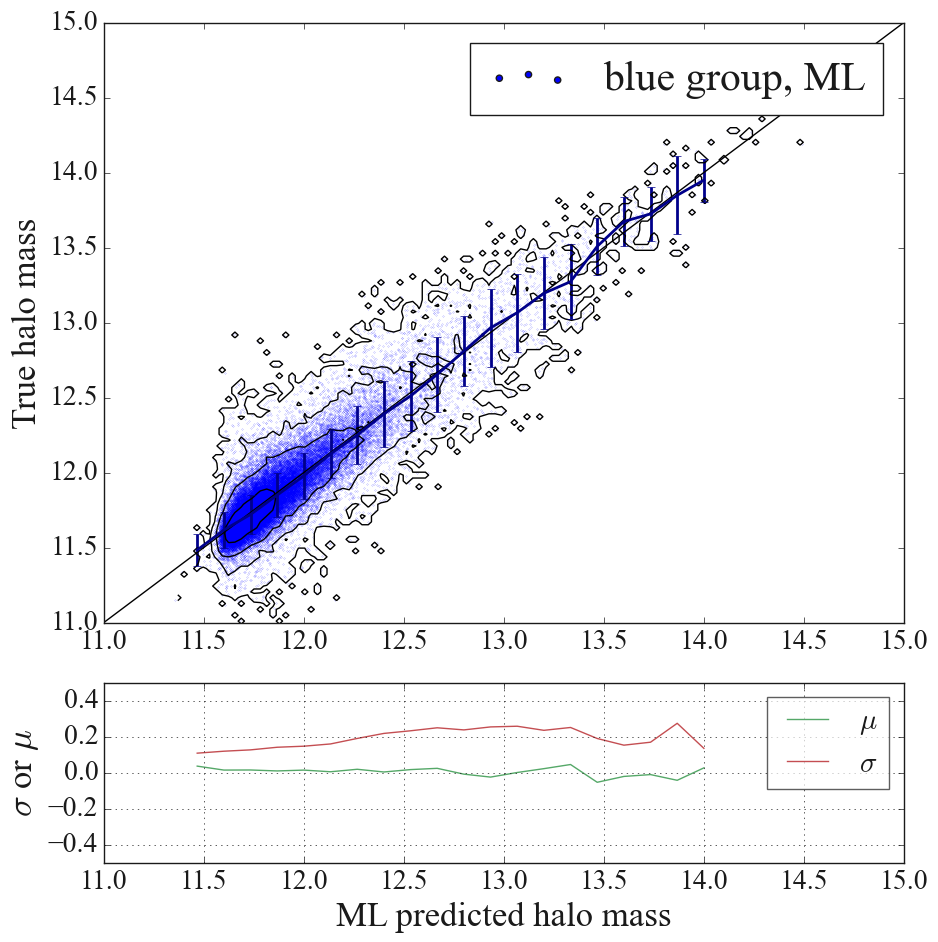}{0.5\textwidth}{}\fig{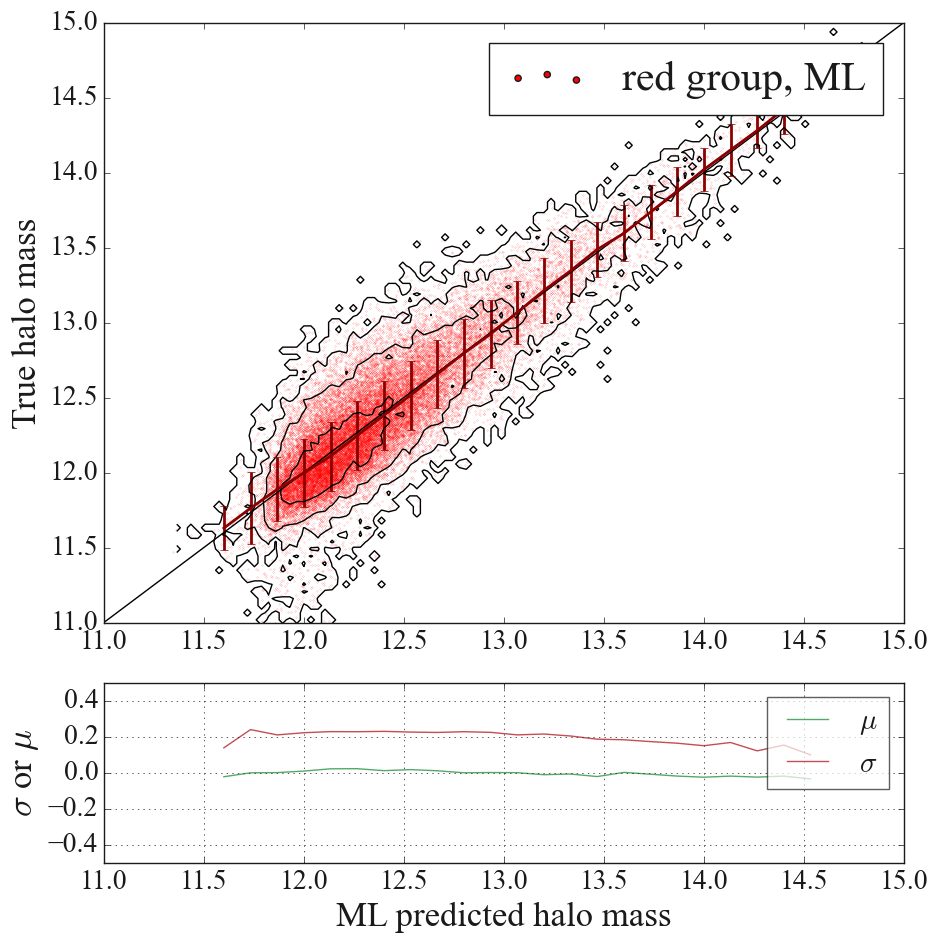}{0.5\textwidth}{}
          }
  \caption{\textbf{Top left panel}: Comparison of AM-predicted halo mass versus true halo mass for blue groups in the complete sample (sample number 1) from the mock catalog, which includes Gaussian noise with $\sigma_n/\sigma_0=1$. Blue dots represent individual DM halos, while the curve with error bars illustrates the mean predicted halo mass within each true halo mass bin; the error bars depict the standard deviation. Black contours indicate the number density, and the black line represents the equality between true and predicted halo mass. In the smaller subpanel, the red and green lines show the mean ($\mu$) and standard deviation ($\sigma$) of the residuals for each predicted halo mass bin, respectively. \textbf{Top right panel}: Same as the top left, but for red groups. \textbf{Bottom left panel}: Same as the top left, but using ML-predicted halo mass for blue groups. \textbf{Bottom right panel}: Same as the bottom left, but for red groups. All masses are measured in units of $\mathrm{M_\odot} h^{-1}$ and presented on a logarithmic scale.}

  \label{fig:AM}
\end{figure*}

\begin{figure*}[t]
  \gridline{\fig{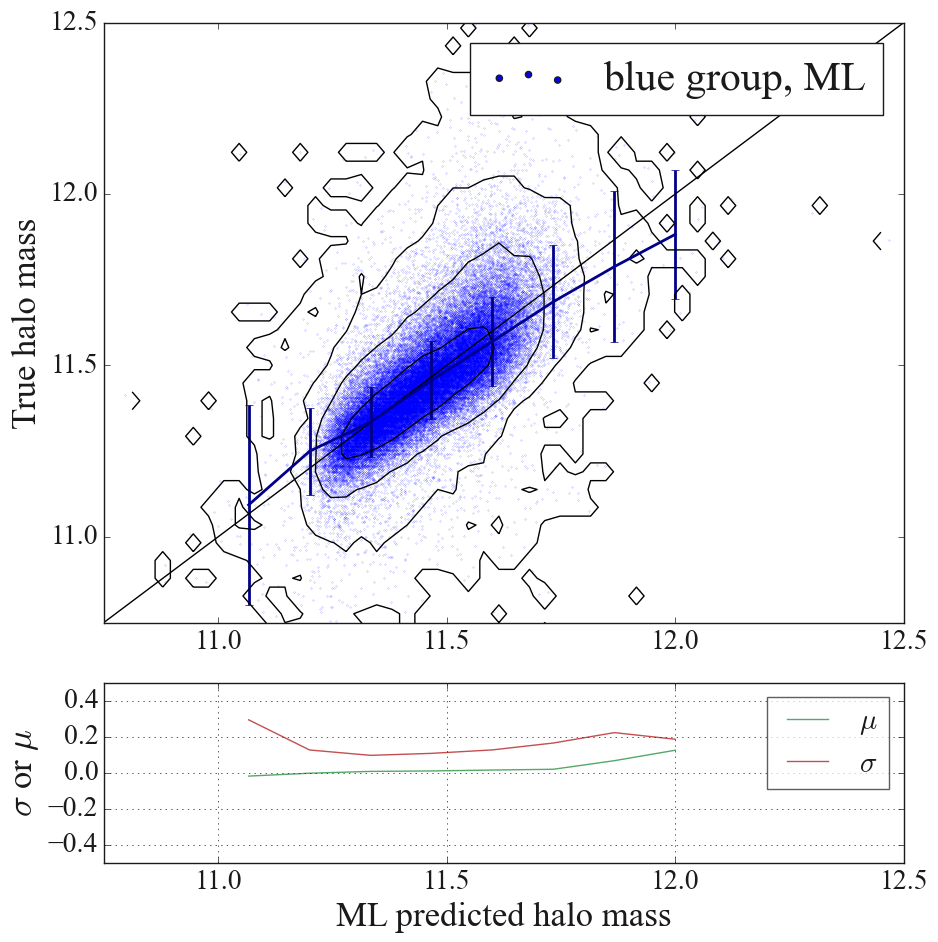}{0.5\textwidth}{}\fig{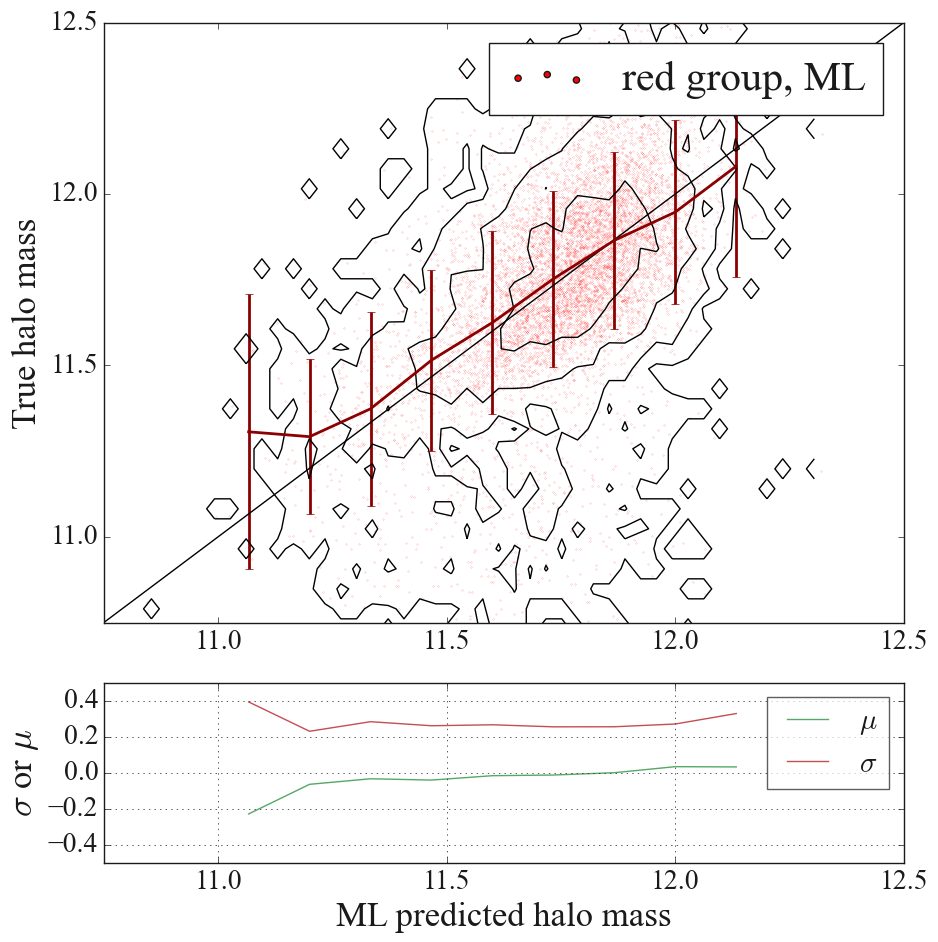}{0.5\textwidth}{}
          }
  \caption{Similar to the top two panels in Figure \ref{fig:AM}, but focusing on the ML-predicted halo mass for blue groups (left panel) and red groups (right panel) within the incomplete sample from the mock catalog.}

  \label{fig:incomplete}
\end{figure*}
\begin{figure*}[t]
  \centering
  \gridline{\fig{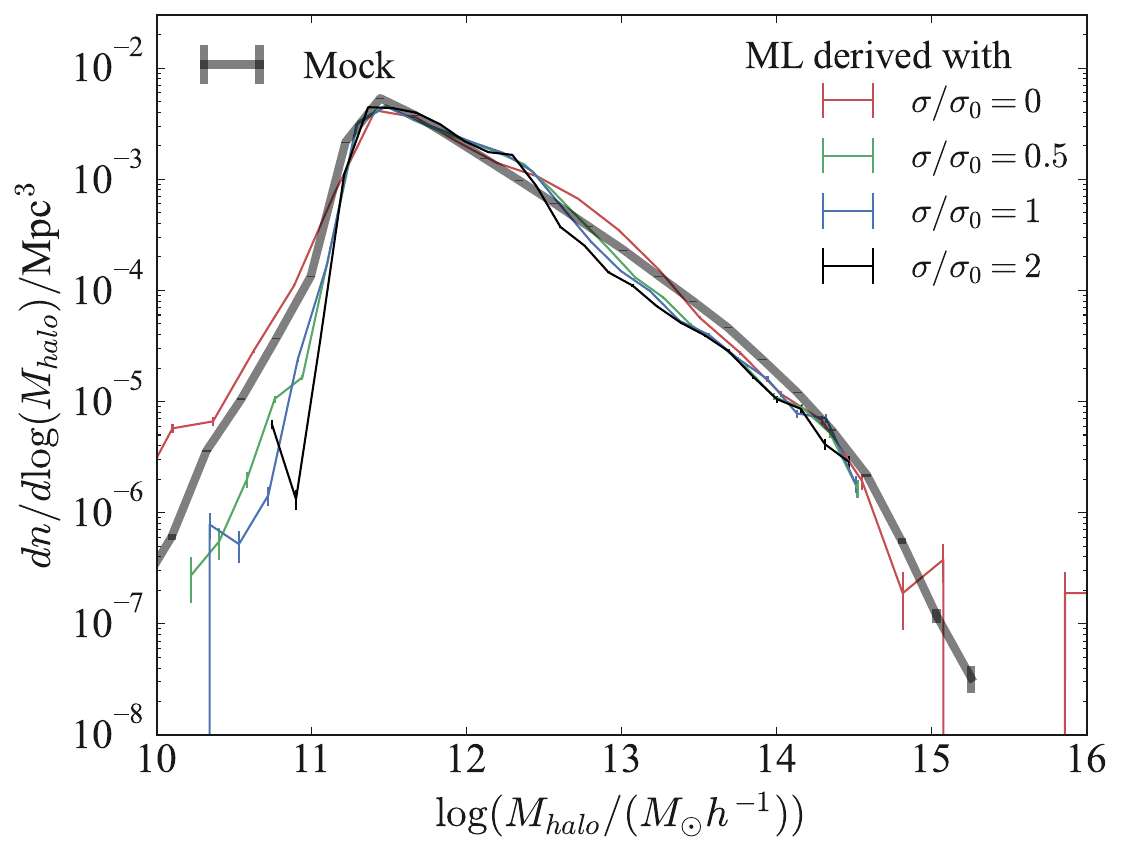}{0.5\textwidth}{}\fig{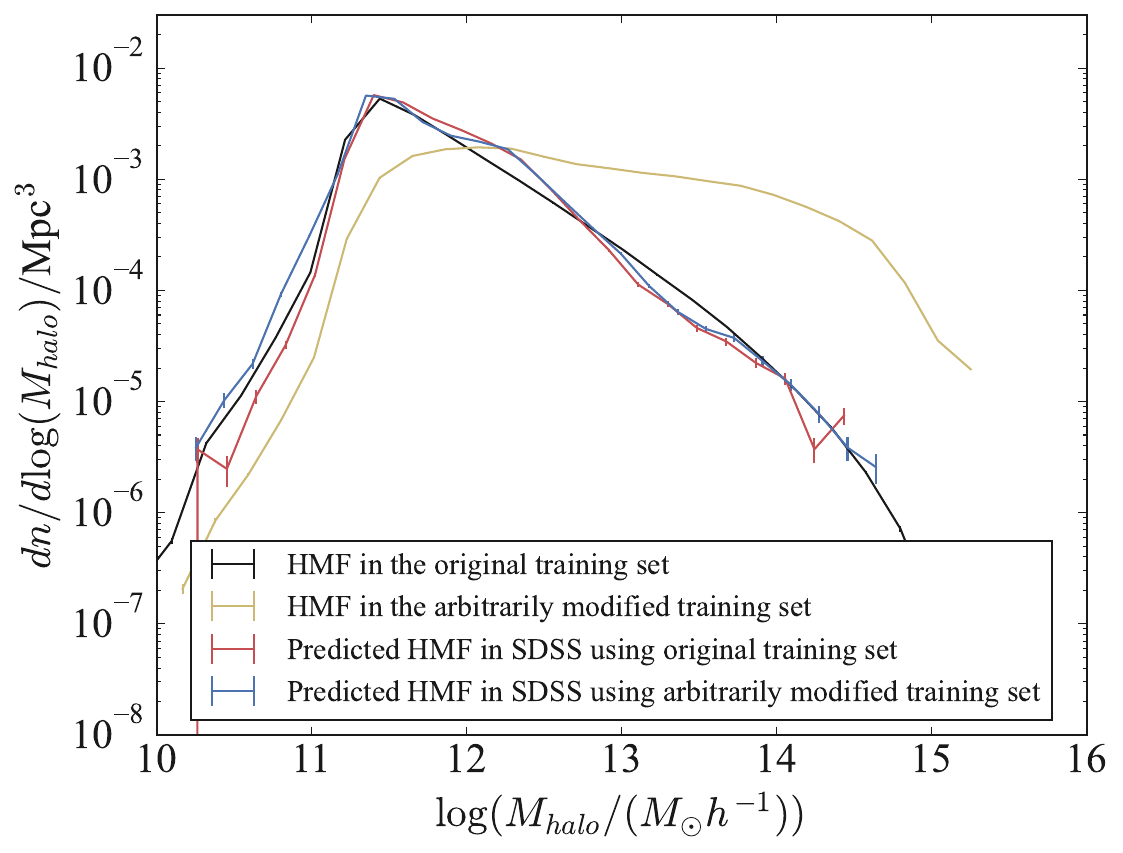}{0.5\textwidth}{}
          }
  \caption{\textbf{Left panel}: HMFs derived from SDSS data using ML models trained on training sets with varying levels of Gaussian noise. The red, green, blue, and black curves represent the HMFs for groups with $0.04<z<0.07$ and $M_{\mathrm{cen}}>10^{9.5}\mathrm{M_\odot}$ in the SDSS data, as predicted by ML models trained on training sets with Gaussian noise levels of $\sigma_n/\sigma_0=$ 0, 0.5, 1, and 2, respectively. The error bars represent the statistical uncertainty in the number of halos within each halo mass bin. All HMFs have been corrected using $V_{max}$ correction. The broad black curve shows the HMF for groups at $z\sim 0.05$ and $M_{\mathrm{cen}}>10^{9.5}\mathrm{M_\odot}$ from the mock catalog. \textbf{Right panel}: The HMFs for groups with $0.01<z<0.04$ and $M_{\mathrm{cen}}>10^{9.5}\mathrm{M_\odot}$ in the SDSS data and the training sets. The black curve represents the HMF in the original training set, while the yellow curve shows the HMF in an arbitrarily modified training set.
  The red curve indicates the HMF obtained by predicting halo masses in the SDSS data using an ML model trained on the original training set, and the blue curve shows the HMF obtained by predicting halo masses in the SDSS data using an ML model trained on the modified training set.}
  \label{fig:HMF}

\end{figure*}
\subsubsection{Sample selection effect}\label{subsubsec:sample_division}

Since the target SDSS spectroscopic sample, upon which the SDSS groups \citep{2007ApJ...671..153Y} are based, is flux-limited, it exhibits a strong selection bias in stellar mass across the sample redshift range, as shown in Figure \ref{fig:samples}. As a consequence, for groups with similar true halo mass, the group richness and total stellar mass (or total luminosity) of observed group members decrease with increasing redshift. Additionally, the observation uncertainty of the parameters utilized varies with redshift. To minimize any potential biases introduced by this sample selection effect across the entire redshift range explored here ($0.01 < z < 0.2$),  we need to process the mock catalog to mimic the selection effect in SDSS observations. The mock catalog provides six snapshots, so we correspondingly split the SDSS data into six redshift ranges based on the luminosity-weighted group center redshifts. Within each redshift range, we set a mass completeness threshold defined as the stellar mass below which the stellar mass function starts to drop. This threshold helps identify galaxies unaffected by sample selection effects, as shown in Figure \ref{fig:samples} and Table \ref{tab:error}. 
This approach allows us to assemble a sizable sample free from selection effect. In the mock catalog, we apply the same mass completeness threshold to segregate galaxies into complete and incomplete samples. However, it is important to note that before this division, we have introduced noise into the mock catalog, see Section \ref{subsubsec:adding_noise}.

The sample selection effect creates a disparity between the SDSS data and the mock catalog. In the mock catalog, we are capable of detecting all existing galaxies, but in the SDSS data, some less luminous (and therefore less massive) galaxies might not be detected. For the complete sample, we assume that all included galaxies are luminous enough to be detected, meaning it does not suffer from the sample selection effect. However, the incomplete sample in the SDSS data suggests that there are additional galaxies that are not detected. This results in a notable difference between the incomplete sample in the SDSS data and its counterpart in the mock catalog. 
To mimic the sample selection effect, we adjust the incomplete sample in the mock catalog to match the $M_\mathrm{cen}$ distribution observed in the SDSS data, as shown in Figure \ref{fig:incom_distribution}.

To avoid the sample selection effect, for both the SDSS data and the mock catalog, when the central galaxy of a group surpasses the mass completeness threshold, we only take the members exceeding the threshold, i.e., all of these galaxies belong to the complete sample, to calculate the properties of the group. Conversely, for a group where its central galaxy falls below the mass threshold, i.e., it belongs to the incomplete sample, we only use the central galaxy to calculate the properties of the group. Therefore, there are just 13 effective properties for these groups, i.e., $M_{\mathrm{cen}}$, SFR, age, $color$.

Through the above approach, we can minimize various potential biases related to redshift, thereby ensuring, in principle, that the predicted halo masses do not contain significant redshift-dependent systematic bias.
\subsubsection{Adding noise and avoiding overfitting}\label{subsubsec:adding_noise}
An important difference between the mock catalog and the SDSS data is the presence of measurement errors in the latter. To make the mock catalog more realistic, we introduce Gaussian noise. Since richness is calculated only for the complete sample, we assume no measurement error and therefore do not add noise.\footnote{The group finder itself can introduce some errors when determining group members. However, since the errors introduced by the group finder are relatively complex and the group catalog used is quite accurate \citep{2007ApJ...671..153Y}, we do not account for these errors in this paper.}  For $color$ and $color_{\mathrm{tot}}$, noise is added to the absolute magnitudes before calculating them. The Gaussian noise introduced to each property has a mean of $\mu_n = 0$ and a standard deviation of $\sigma_n = \sigma_0 = (P84-P16)/2$, as detailed in Section \ref{subsec:group-catalog} and Table \ref{tab:error}.

As star-forming and passive galaxies have different measurement errors, we define a boundary in the mock catalog to distinguish them using Equation \ref{eq:rb_mock_first}: 
\begin{equation}\label{eq:rb_mock_first}
    \log(\mathrm{SFR})=0.7\log(M_*)-7.7
\end{equation}
where SFR is measured in $\mathrm{M_\odot/yr}$ and stellar mass ($M_{*}$) in $\mathrm{M_\odot}$. Additionally, to evaluate our results across varying noise levels, we also add Gaussian noise with $\sigma_n = 0, \sigma_n = 0.5\sigma_0,$ and $\sigma_n = 2\sigma_0$ to the mock catalog. After introducing these noises, to maintain consistency with the observational data, we exclude galaxies in the mock catalog with stellar masses below $10^{9.5}\mathrm{M_\odot}$.

Another discrepancy lies in the SFRs of fullly quenched galaxies with very low or zero SFR. In the SDSS data, these SFRs cannot be measured satisfactorily and most of them are just upper limits, whereas the SFRs of these galaxies in the mock catalog are accurately determined. This results in the SFR distribution in the mock catalog being different from that in the SDSS data. Therefore, we cannot directly use the original SFR of these galaxies in the mock catalog. Instead, we reassign the SFR of the galaxies in the mock catalog that fall below Equation \ref{eq:ex_threshold} using Equation \ref{eq:new_sfr} to mimic the observed SFR distribution. 

\begin{equation}\label{eq:ex_threshold}
  \log(\mathrm{SFR}) = (0.7+0.02i)\log(M_{*})-8.4-0.02i
\end{equation}
\begin{equation}\label{eq:new_sfr}
  \log(\mathrm{SFR})=(0.7+0.02i)\log(M_{*})-8.65-0.02i+\mathcal N(0,0.27)
\end{equation}
Here, $i$ represents the sample number, SFR is measured in $\mathrm{M_\odot/yr}$, $M_{*}$ in $\mathrm{M_\odot}$. $\mathcal N(0,0.27)$ indicates a Gaussian noise with a mean of zero and a standard deviation of 0.27. These adjustments ensure that the SFR distribution of passive central galaxies in the mock catalog closely mirrors that of the SDSS data, as shown in Figure \ref{fig:com_distribution_red} and bottom panel of Figure \ref{fig:incom_distribution}.

Another advantage of introducing noise into the mock catalog is that it can mitigate the risk of overfitting in the ML model, see Section \ref{sec:noise_importance} and Figure \ref{fig:noise_com}.

\subsubsection{Data reprocessing and normalization}\label{subsubsec:normalization}
Given the different SHMRs between red and blue groups, as discussed in Section \ref{sec:intro}, we categorize the groups into red and blue. Groups with passive central galaxies are classified as red groups, while those with star-forming central galaxies are classified as blue groups. This classification not only aligns with the inherent characteristics of the galaxies but also facilitates the normalization process, ensuring that the distribution of galaxy properties within each group category tends toward a single-peak distribution. Due to the bimodal distributions in the SFR-stellar mass plane, which vary across different redshifts, we employ slightly different boundaries for each redshift range to effectively separate these distributions. The boundaries for the SDSS data and the mock catalog across all redshift ranges are defined by Equation \ref{eq:rb_obs} and Equation \ref{eq:rb_mock}, respectively.

\begin{equation}\label{eq:rb_obs}
  \log(\mathrm{SFR})=(0.6+0.02i)\log(M_{\mathrm{cen}})-6.9-0.12i
\end{equation}
\begin{equation}\label{eq:rb_mock}
  \log(\mathrm{SFR})=(0.7+0.02i)\log(M_{\mathrm{cen}})-7.9+0.03i
\end{equation}

Here, SFR is measured in $\mathrm{M_\odot/yr}$, $M_{\mathrm{cen}}$ in $\mathrm{M_\odot}$ and $i = 0, 1, 2, 3, 4, 5$ represents the sample number corresponding to different redshift ranges detailed in Table \ref{tab:error}. By utilizing these boundaries, we ensure that, within each redshift bin and corresponding snapshot, the distributions of SFR for both star-forming and passive central galaxies in the mock catalog closely resemble those in the SDSS data.

For any simulation, the distribution of absolute values of physical quantities cannot fully match the observed data; in some cases, they can show significant discrepancies.
To minimize the impact from these differences,
for both the SDSS data and the mock catalog, we implement normalization in each subsample, similar as done in \cite{2019ApJ...881...74M}:
\begin{equation}\label{eq:nor}
  P_n = \frac{P-\mu_P}{\sigma_P}
\end{equation}
Here, $P$ represents SFR, age, $color$, and $color_{\mathrm{tot}}$. The mean ($\mu_P$) and standard deviation ($\sigma_P$) of each property's distribution are used to scale $P$. To mitigate the impact of outliers on the normalization process, we exclude values more than $6\sigma_P$ away from $\mu_P$. Subsequently, we recalculate $\mu_P$ and $\sigma_P$ using this refined dataset and apply these parameters to Equation \ref{eq:nor}. 

As an example, we present the distribution of each calibrated property for groups within the redshift range $0.04<z<0.07$ in Figure \ref{fig:com_distribution_blue}, Figure \ref{fig:com_distribution_red}, and Figure \ref{fig:incom_distribution}. These figures respectively correspond to the blue groups in the complete sample, the red groups in the complete sample, and both blue and red groups in the incomplete sample. For the mock catalog, we have introduced Gaussian noise with $\sigma_n=\sigma_0$. These figures demonstrate that, across each property, the distributions closely align between the mock catalog and the SDSS data, particularly for the blue groups.

In summary, we have organized the groups into six redshift bins, classified them into complete or incomplete samples, applied four kinds of noise to the mock catalog, and categorized them into red or blue groups. This process results in 88 subsamples from the mock catalog and 22 subsamples from the observational data.

\subsection{Machine Learning}\label{subsec:ML}
We utilize \texttt{XGBoost} \citep{2016arXiv160302754C} as the ML model. It is an efficient, powerful, and easy-to-use implementation of the gradient-boosted tree algorithm. For each subsample, we construct an ML model to predict the DM halo mass. The features of the ML model depend on whether it is trained on a subsample from the complete or incomplete sample. For the complete sample, 25 properties are utilized to characterize the groups, while the incomplete sample groups are described using 13 properties, as shown in Section \ref{subsec:group-property}. 

Each subsample within the mock catalog is divided into three parts according to an 8:1:1 ratio to form the training set, validation set, and test set, respectively. To identify the ML model that yields the best predictions, we employ the Python package \texttt{hyperopt}\footnote{\url{https://github.com/hyperopt/hyperopt}} to search the hyperparameter space constructed by \texttt{max\_depth}, \texttt{n\_estimators}, \texttt{learning\_rate}, \texttt{subsample}, and \texttt{min\_child\_weight}. Surprisngly, we discover that the hyperparameters have minimal impact on the predictive outcomes. Consequently, for efficiency, we train our models with \texttt{n\_estimators} set to 100 and other hyperparameters at their default settings. This approach not only saves time but, as evidenced by our tests, also delivers models that perform nearly optimally.

For the SDSS data, we apply the trained models to estimate halo masses of galaxy groups. However, approximately 11\% of central galaxies in the SDSS data lack age measurements. To avoid the impact of the missing data, we train additional models using the existing training set but exclude age as a feature. The performance of these new models, which do not include age, is similar to the original models that incorporate age as a feature. The results are detailed in Table \ref{table:complete_error} for the complete sample and Table \ref{table:incomplete_error} for the incomplete sample.

\section{Performance of ML method in test set}\label{sec:test_set}
\subsection{Which noise level assigned on halo mass that performs the best?}\label{sec:noise_importance}
In this study, we trained various ML models using training sets with different levels of Gaussian noise for halo mass prediction. To elucidate which model performs optimally, we train several models using training sets that incorporate varying levels of noise, and then utilize these models to make predictions on test sets with different noise levels. By assessing the accuracy of these models on identical test sets, we determine which model is the most effective. Figure \ref{fig:noise_com} illustrates the results. 
This figure shows that for a test set with a certain level of noise, the accuracy of the ML model is highest only when the noise level in the training set matches that in the test set, as marked by the stars.

Thus, for practical applications to the SDSS data, we recommend using a model trained with a noise level matching the observational noise level, ideally $\sigma_n/\sigma_0 = 1$. However, since Gaussian noise is simply added to the training set, there's a risk that the measurement error could be underestimated or overestimated for certain galaxy populations. Consequently, for samples where the average measurement error differs from $\sigma_0$, models trained with $\sigma_n/\sigma_0 = 0.5$ or $2$ might yield more accurate halo mass estimates.

The behavior described above, where the accuracy of the ML model is highest only when the noise level in the training set matches that in the test set, is analogous to the phenomena of overfitting and underfitting in ML. An optimal ML model benefits from a training set with minimal noise, as excessive noise can obscure valuable information inherent in the data. When a model trained on a less noisy dataset is applied to a noisier test set, it may overinterpret the data or utilize irrelevant information, resulting in poor performance akin to overfitting. Conversely, a model trained on a noisier dataset may lack sufficient information when applied to a cleaner test set, leading to inadequate predictions reminiscent of underfitting. The most effective model typically comes from matching the noise levels of the training and test sets, ensuring that the model captures the right amount of information. By carefully adjusting the noise in the training set to reflect that of the test set, we can prevent overfitting and enhance the model’s accuracy and robustness.

\subsection{Estimation error in the test set}
In this section, we evaluate the estimation error of the best-performing ML models on the test set, where the training and test sets have equivalent noise levels. We assess the accuracy of the ML method by calculating the mean and standard deviation of the residuals of halo mass estimations for the complete and incomplete samples, as presented in Table \ref{table:complete_error} and Table \ref{table:incomplete_error}, respectively. The error of the AM method on the complete sample is also shown in Table \ref{table:complete_error} as a comparison. Table \ref{table:complete_error} demonstrates that the mean of residuals for the ML method across all subsamples is very close to zero, unlike the AM method. The AM method tends to overestimate the halo mass of blue groups and underestimate that of red groups, indicating a bias. In contrast, the ML method offers an unbiased estimation of halo mass across various galaxy populations with different star-forming states, making a significant improvement over the AM method. Furthermore, the standard deviation of residuals for the ML method is $\sim 1/3$ smaller than that derived from the AM method, underscoring the enhanced precision of the ML approach.

To further clarify the results obtained from the ML method, we show Figure \ref{fig:AM}. This figure illustrates the complete sample with sample number = 1 in the mock catalog, to which we have added Gaussian noise with $\sigma_n=\sigma_0$. 
The top two panels demonstrate that the AM method tends to underestimate the halo masses of red groups and overestimate them for blue groups, particularly overestimating the halo masses of blue groups with $M_{\mathrm{halo}} \gtrsim 10^{11.7}\mathrm{M_\odot} h^{-1}$ and underestimating those of red groups with $M_{\mathrm{halo}} \lesssim 10^{12.5}\mathrm{M_\odot} h^{-1}$. This effect will lead to a significant bias in their relative halo masses around $\log(M_{\mathrm{halo}}/\mathrm{M_\odot})\sim12-12.5$ , where there are comparable numbers of red and blue groups.
Conversely, the bottom two panels show that the ML method provides an unbiased estimation of halo masses for both blue and red groups across all mass bins, evidenced by the green curves in these subpanels being very close to zero. Furthermore, the lower standard deviation of the residuals from the ML method compared to the AM method is evident as the red curves in the subpanels of the top two panels are consistently higher than those in the bottom panels, highlighting the superior precision of the ML method.

Additionally, we present the results of the ML-predicted halo mass versus true halo mass for the incomplete subsample with sample number 1 in Figure \ref{fig:incomplete}. In this instance, we omit the comparison with the AM method, as it generally performs poorly in predicting low-mass halos.  This limitation is noted in practices such as \cite{2007ApJ...671..153Y}, where the AM method is only used to estimate the masses of halos greater than $10^{11.6}\mathrm{M_\odot} h^{-1}$. In contrast, the ML method demonstrates reliable accuracy for halos less massive than $10^{11.3}\mathrm{M_\odot} h^{-1}$ or even lower and provides unbiased estimates of halo mass across all mass ranges. Thus, it proves effective in low-mass regions traditionally challenging for the AM method, offering precise halo mass estimations.

To further investigate the variation in prediction errors with halo mass, the trends observed in the smaller subpanels of Figures \ref{fig:AM} and \ref{fig:incomplete} reveal distinct behaviors for blue and red groups. For blue groups, the prediction error decreases with lower halo mass, indicating more accurate estimates for smaller halos. This trend can be attributed to the simpler halo assembly histories and SFH at the low-mass end, which facilitate more precise predictions based on group member properties. Similar discussions are also found in \cite{2024ApJ...972..108L}.

In contrast, this trend is not observed for red groups; in fact, the prediction errors tend to be smaller at the high-mass end ($M_\mathrm{halo} > 10^{13.5}M_\odot/h$). 
This may be due to the fact that, although low-mass halos generally have simpler assembly histories, quenching processes can decouple the growth of the halo from that of the galaxy, complicating the mapping from group properties to halo mass. Stellar age serves as an indicator of quenching time, but its observational uncertainty is considerable, especially for quenched galaxies, making it challenging to accurately constrain the stellar-to-halo mass ratio for red groups based on stellar age. While higher-mass halos do indeed have more complex assembly histories, the richness parameter becomes a significant factor at this scale. In most low-mass groups, richness is typically 1 and provides little information. However, for massive halos, richness carries substantial information regarding the merger history, thereby enhancing the precision of halo mass estimates (e.g., \citealp{2012ApJ...746..178R}).

\section{RESULTS} \label{sec:results}

\subsection{Halo mass function in observation}\label{subsec:hmf}
The halo mass function (HMF) shows the distribution of halo masses and provides an indirect statistical check on the accuracy of halo mass estimates. If the estimated halo masses do not reproduce the actual HMF, this suggests inaccuracies in the estimation process. The left panel of Figure \ref{fig:HMF} displays the HMFs from the SDSS data and the mock catalog. This panel shows the HMFs from SDSS data in the redshift bin $0.04<z<0.07$, derived from ML models trained on training sets with Gaussian noise levels of $\sigma_n/\sigma_0=$ 0, 0.5, 1, and 2, respectively. It also displays the HMF of galaxy groups with $M_{\mathrm{cen}}>10^{9.5}\mathrm{M_\odot}$ at $z\sim0.05$ from the mock catalog. This panel shows that the ML-derived HMFs closely match the HMF in the mock catalog. The observed differences between the ML-derived HMFs, which exceed the statistical errors, can be attributed to variations in the training sets. HMFs are obtained from a model trained on a dataset with a different level of added noise. As discussed in Section \ref{sec:noise_importance}, the model that shows the best agreement with observational data is the one trained with noise levels matching the observational uncertainties. Consequently, it is expected that models trained with different noise levels would produce HMFs that deviate from this optimal model, thereby reflecting systematic uncertainties in the predictions.

Additionally, we show the HMFs in different redshift bins in Figure \ref{fig:hmf_redshift}. The consistency of the HMFs at the high-mass end indicates that our halo mass estimates do not exhibit significant redshift-dependent systematic bias. The differences observed at the low-mass end allow us to determine the completeness limit of the estimated halo masses. For example, in the redshift bin [0.04,0.07], the halo mass estimates are complete for masses above $10^{11.5}M_\odot$. For further details, please refer to Section \ref{sec:hmf_redshift}.


Notably, the AM method can reproduce the real HMF because it directly uses the actual HMF as input, allowing it to replicate the distribution exactly by definition. In contrast, the ML method estimates halo masses individually without any prior HMF knowledge, then aggregates these estimates to derive the observed HMF. This process underscores the accuracy of the ML method, as the derived HMF closely mirrors the actual distribution without relying on predefined inputs. 

It might be argued that the HMF is inherently represented within the training set.
To clarify, within the redshift range $0.01<z<0.04$, we constructe an arbitrarily modified training set by resampling the original training set. We then use this modified training set to train an ML model, which is subsequently applied to predict halo masses in the SDSS data, thereby obtaining the HMF. We show the result in the right panel of Figure \ref{fig:HMF}. There is a significant difference between the HMF of the original training set and that of the arbitrarily modified training set. However, the HMFs in the SDSS data, whether based on the original or modified training set, are very similar.
Therefore, the HMF obtained by ML method does not merely replicate the training set’s HMF but rather reflects a genuinely independent outcome, further validating the ML method's efficacy in accurately estimating halo masses across varied mass ranges.

\subsection{Compare with weak lensing measurements}\label{subsec:weak_lensing}

\begin{figure*}[t]
  \centering
  \gridline{\fig{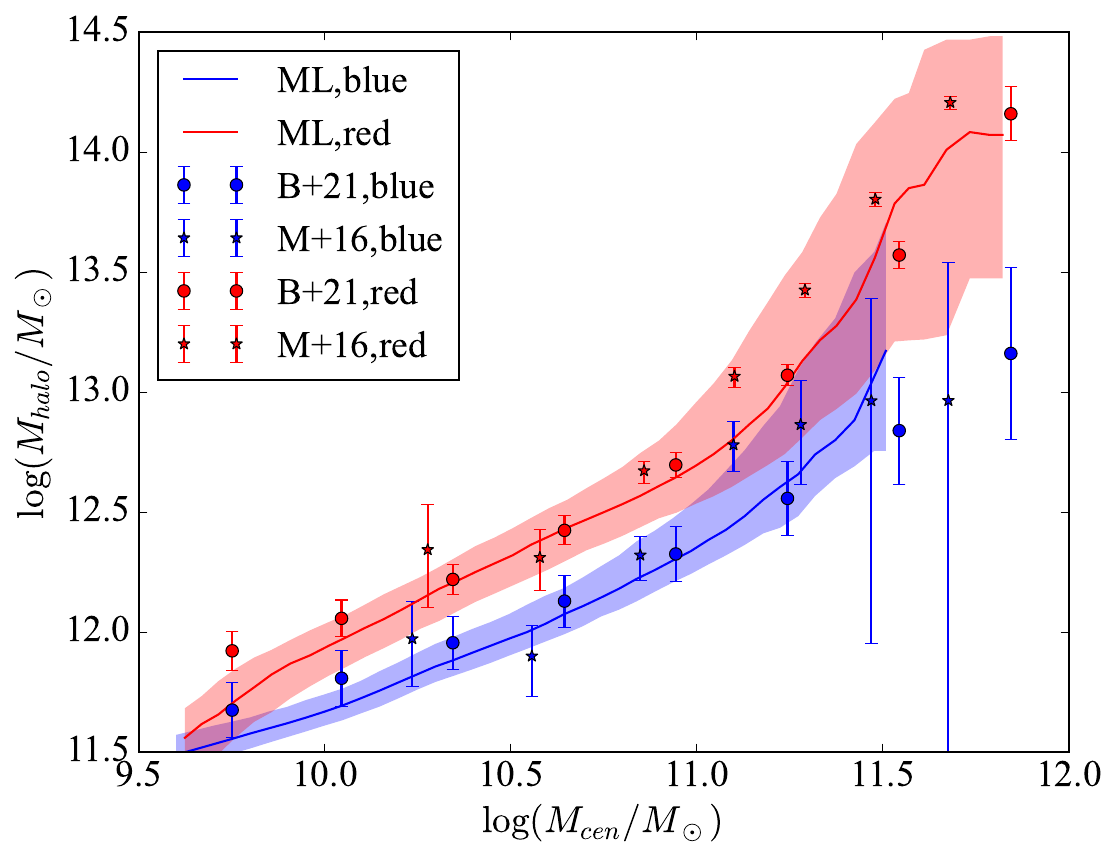}{0.5\textwidth}{}
            \fig{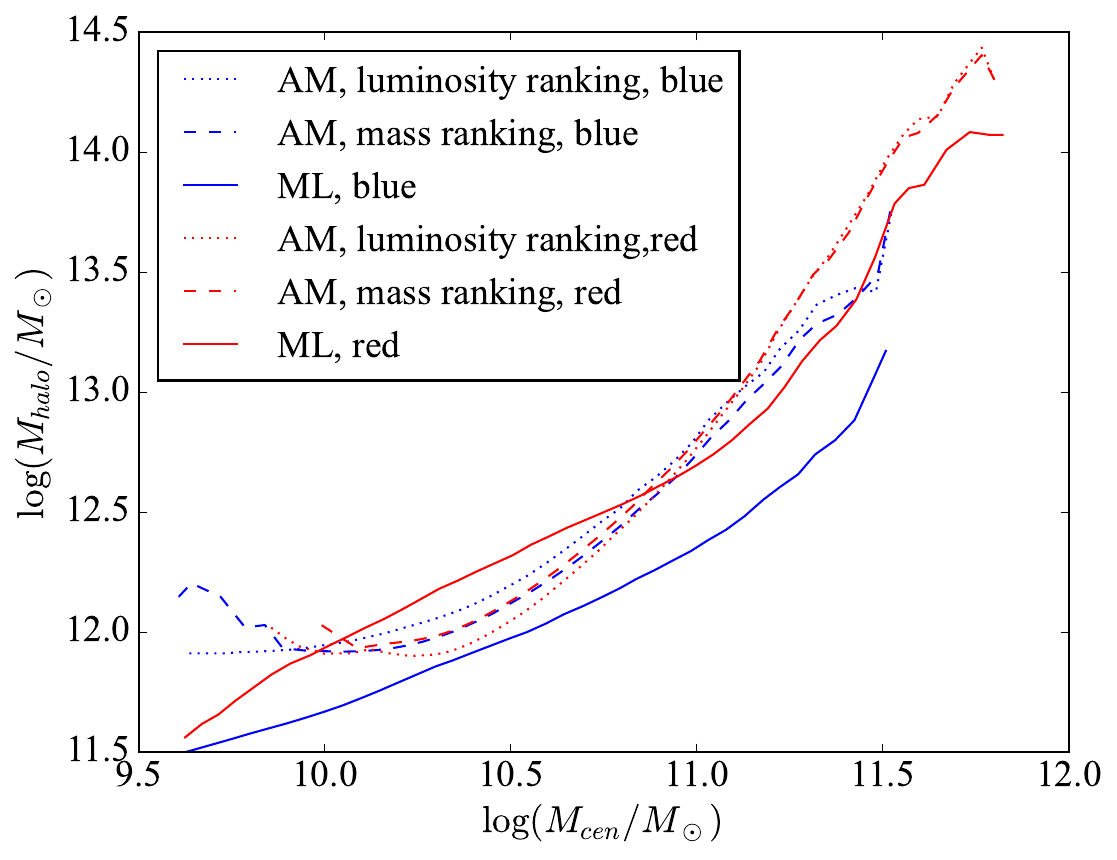}{0.5\textwidth}{}
          }
  \gridline{\fig{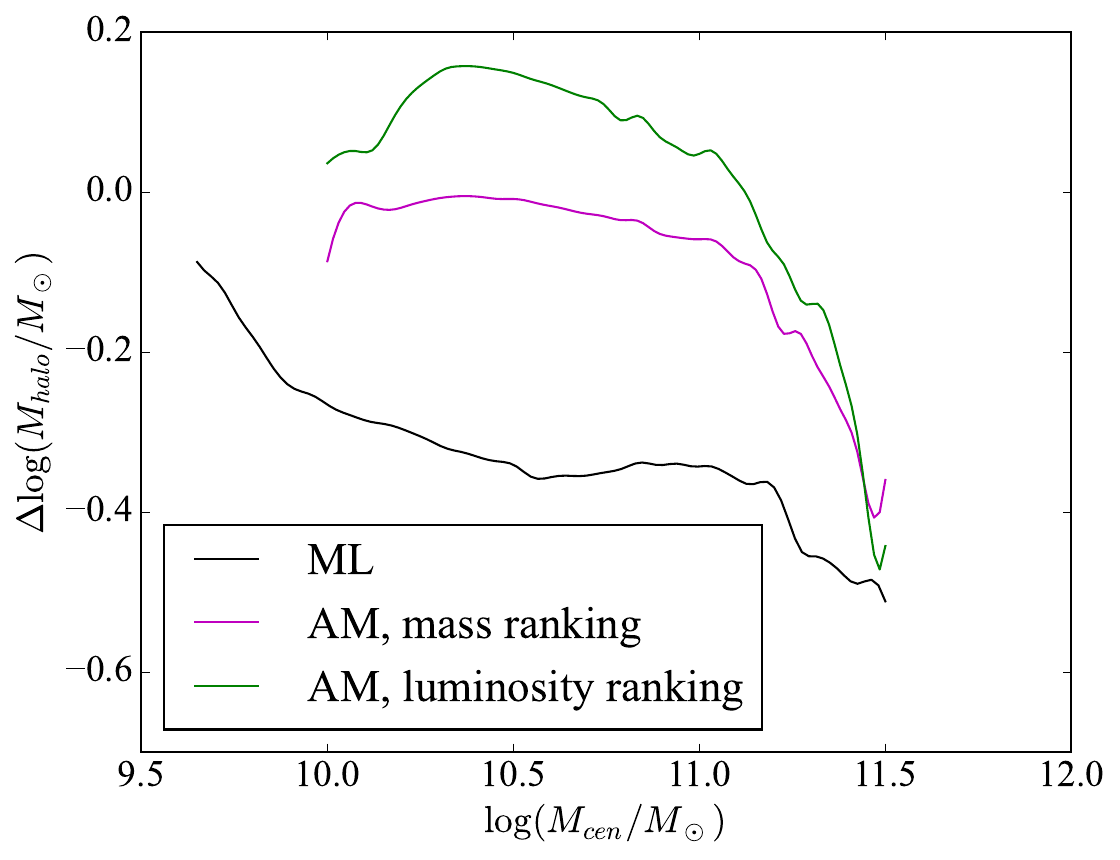}{0.5\textwidth}{}
            \fig{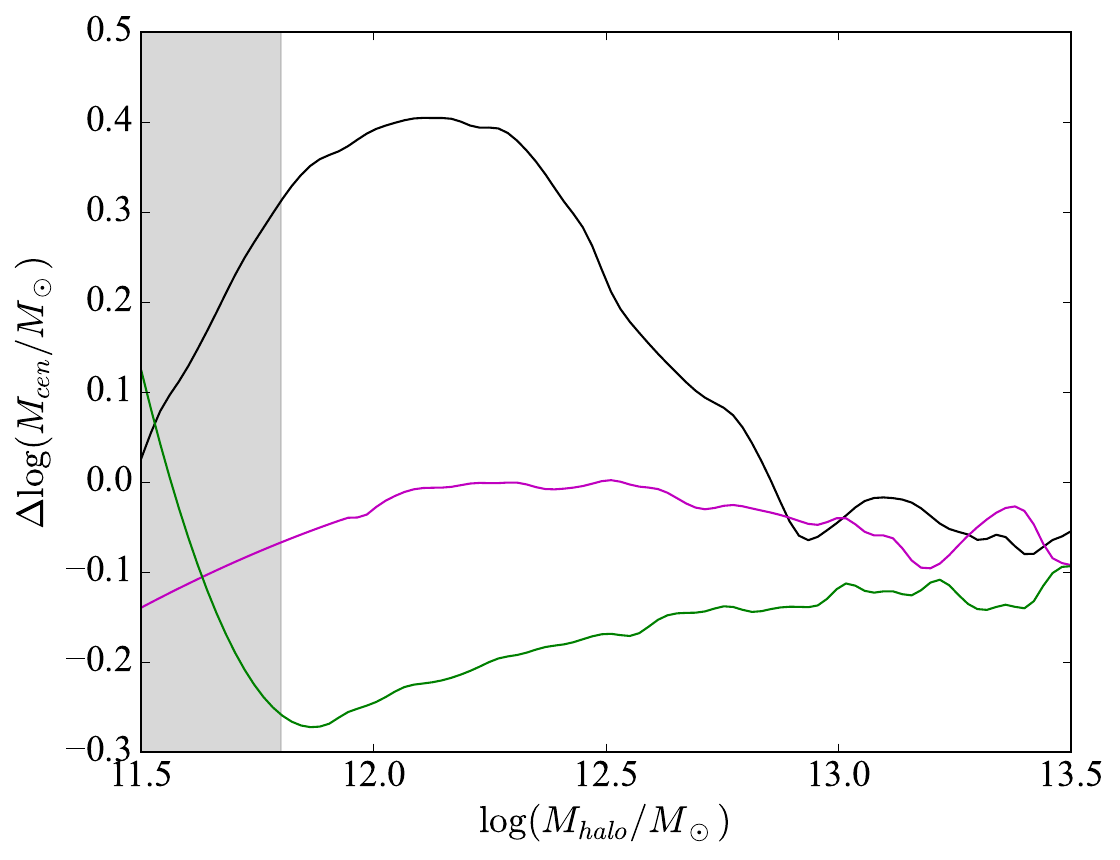}{0.5\textwidth}{}
          }
  \caption{\textbf{Top left panel}: The solid curves, dots, and stars in blue/red illustrate the SHMRs of blue/red groups as derived from the ML method, the weak lensing measurements from \cite{2016MNRAS.457.3200M}, and \cite{2021A&A...653A..82B}, respectively. The shaded regions and error bars indicate the 16th to 84th percentile ranges. \textbf{Top right panel}: The solid, dashed, and dotted curves in blue/red depict the SHMRs of blue/red groups obtained from the ML method, and the AM method according to total stellar mass ranking and total $r$-band luminosity ranking from \cite{2007ApJ...671..153Y}, respectively. The solid curves in this panel are identical to those in the left panel. 
  \textbf{Bottom left panel}: The difference in halo mass between blue and red groups, as derived by the ML method (black), the AM method based on mass ranking (magenta), and the AM method based on luminosity ranking (green), is shown as a function of central stellar mass. \textbf{Bottom right panel}: The difference in central stellar mass between blue and red groups is shown as a function of halo mass, as derived by the ML method (black), the AM method based on mass ranking (magenta), and the AM method based on luminosity ranking (green). The shaded region indicates where $\log(M_\mathrm{halo}/M_\odot)<11.8$, corresponding to the halo mass range where our sample becomes incomplete, as shown in the left panel of Figure \ref{fig:HMF}. The SHMRs derived from both the ML method and the AM method are based on groups within the redshift range $0.04<z<0.07$. For the ML method, halo masses are obtained from a model trained on a dataset with Gaussian noise characterized by $\sigma_n/\sigma_0=1$.}

  \label{fig:weak_lensing}
\end{figure*}

The SHMR serves as an additional statistical examination of the halo mass estimation. While the HMF outlines the overall distribution, the SHMR establishes a scaling relationship between the $M_{\mathrm{cen}}$ of groups and their corresponding host halo mass. In Section \ref{subsec:hmf}, the ML-derived HMFs align closely with those in the mock catalog and literature. Therefore, a correct SHMR, in conjunction with the HMF, will underpin the reliability of the ML-derived halo masses. In this section, we compare the SHMRs derived from halo masses estimated by the ML method, the AM method, and weak lensing measurements. As mentioned in Section \ref{sec:intro}, at a given stellar mass, the red groups locate in more massive DM halos than blue groups. Top two panels of Figure \ref{fig:weak_lensing} presents the SHMRs of blue and red groups separately. 

In the top left panel of Figure \ref{fig:weak_lensing}, we present the SHMRs of blue and red groups as determined by our ML method and the weak lensing measurements reported by \cite{2016MNRAS.457.3200M} and \cite{2021A&A...653A..82B}, respectively. The SHMRs derived from the ML method are based on groups within the redshift range $0.04<z<0.07$, with halo masses obtained from the ML model trained on training set with Gaussian noise characterized by $\sigma_n/\sigma_0=1$. It is important to note that our study distinguishes between blue and red group based on their central galaxy's SFR, whereas the weak lensing studies classify them by color. The analysis reveals that the ML-derived SHMRs closely match the weak lensing results, demonstrating that at a fixed stellar mass, red groups reside in more massive halos than blue groups. Specifically, for red groups, the ML-derived SHMR aligns remarkably well with the result from \cite{2021A&A...653A..82B}. Although the SHMR from \cite{2016MNRAS.457.3200M} shows slightly higher halo mass at the same stellar mass, it still falls within the 16th to 84th percentile range of the ML-derived SHMR. This offset may result from differences in sample selection. \cite{2016MNRAS.457.3200M} derived their SHMRs using Locally Brightest Galaxies, defined as the $r$-band brightest galaxies within 1 projected Mpc and 1000 km/s. Similarly, for blue groups, the ML-derived SHMR also agrees with those derived from weak lensing measurements.

To more clearly illustrate the differences in SHMR between the blue and red groups, we show the difference in halo mass for a given central stellar mass in the bottom left panel of Figure \ref{fig:weak_lensing}. The bottom right panel presents the difference in central stellar mass for a given halo mass.
In the bottom left panel, as central stellar mass increases, the difference in halo mass gradually increases, ranging from -0.1 dex at $\log(M_\mathrm{cen}/M_\odot) = 9.5$ to -0.5 dex at $\log(M_\mathrm{cen}/M_\odot) = 11.5$. In the bottom right panel, the maximum difference in central stellar mass is 0.4 dex, occurring at $\log(M_\mathrm{halo}/M_\odot) \sim 12.0$, where the number of blue and red groups is similar, and which also corresponds to the location of the golden mass.

In the top right panel of Figure \ref{fig:weak_lensing}, we also show the SHMRs derived by the AM method
based on total stellar mass ranking and total $r$-band luminosity ranking, respectively.
This panel shows that the SHMRs from the AM method based on total stellar mass ranking are quite similar for both blue and red groups, especially at $M_{\mathrm{cen}}\sim10^{10.5}\mathrm{M_\odot}$. However, the SHMRs derived from the AM method based on $r$-band luminosity demonstrate a trend contrary to those obtained from the ML method or weak lensing measurements, suggesting that at $M_{\mathrm{cen}}\sim 10^{10.5}\mathrm{M_\odot}$, blue groups reside in more massive halos than red groups. Notably, in \cite{2007ApJ...671..153Y}, halo less massive than $10^{11.6}\mathrm{M_\odot}/h$ are excluded, so the SHMRs are unreliable for $M_{\mathrm{cen}}\lesssim 10^{10.0}\mathrm{M_\odot}$. The discrepancies between AM-derived SHMRs and those measured by weak lensing indicate that the AM method tends to overestimate halo masses for blue groups and underestimate those for red groups at $M_{\mathrm{cen}}\sim10^{10.5}\mathrm{M_\odot}$ and $M_{\mathrm{halo}}\sim10^{12}\mathrm{M_\odot}$. These biases are consistent with Figure \ref{fig:AM}. 

Combining the results from the HMF and SHMR, we can conclude that our ML method provides accurate halo mass estimations. Furthermore, this suggests that the foundation of our prediction, i.e., the \cite{2007ApJ...671..153Y} SDSS group catalog, is reliable and the sample incompleteness has been properly corrected. If this were not the case, the HMF would deviate from the theoretical expectations.

\begin{figure*}
  \centering
  \gridline{\fig{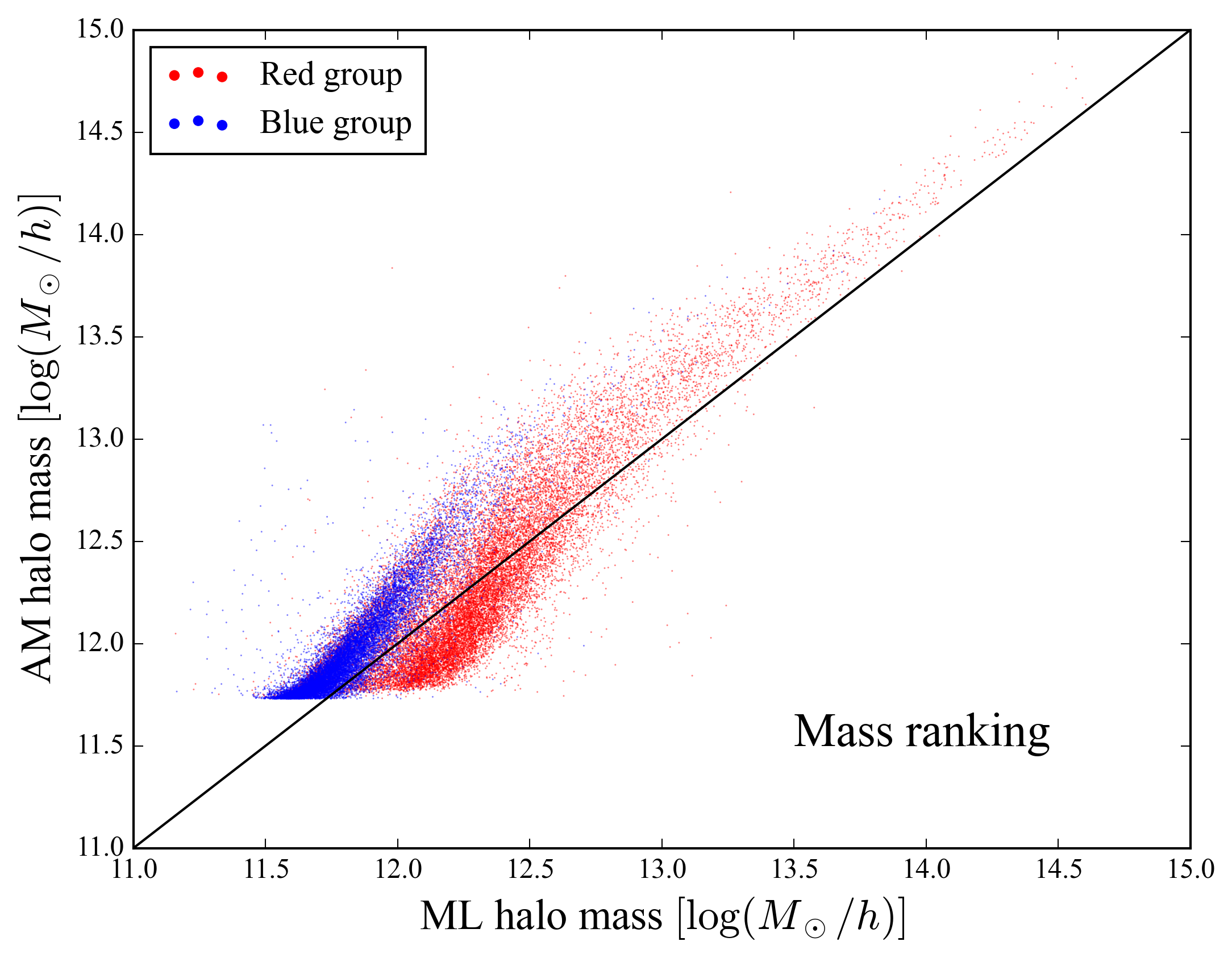}{0.5\textwidth}{}
            \fig{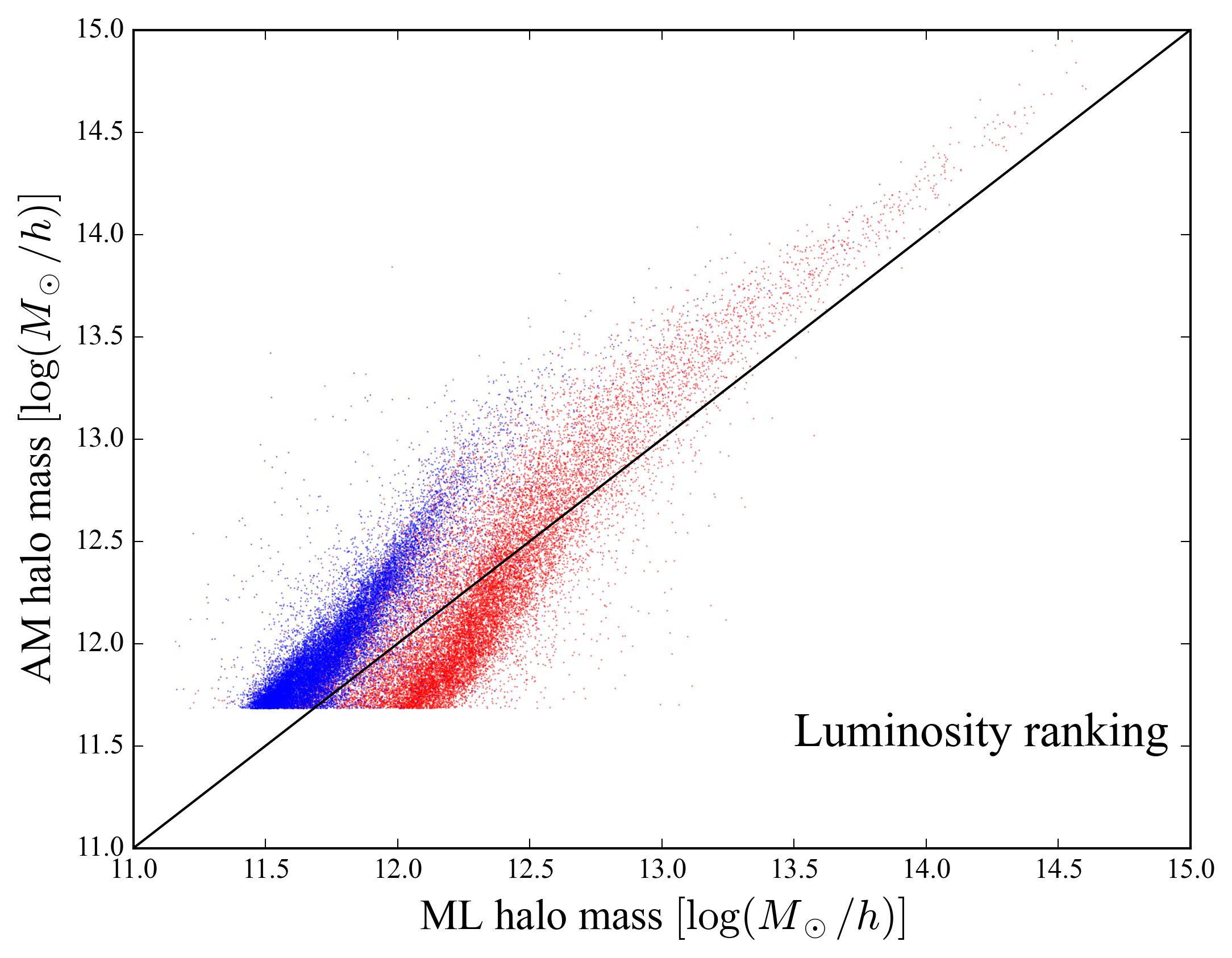}{0.5\textwidth}{}
          }
  \caption{Comparison between the halo masses from our ML method and those from \cite{2007ApJ...671..153Y}. The red and blue points indicate the red and blue groups, respectively. The black line indicates where the two estimates are equal. The left panel shows the comparison based on total stellar mass ranking, while the right panel is based on total $r$-band luminosity ranking. 
  }
  \label{fig:comparison}
\end{figure*}

\subsection{Comparison with halo mass derived by Abundance Matching}\label{subsec:yang}

In this section, we compare the halo masses derived from our ML method with those from the \cite{2007ApJ...671..153Y} catalog, which uses abundance matching method based on total stellar mass and total $r$-band luminosity ranking.

As shown in Figure \ref{fig:comparison}, the halo masses derived from the ML method are generally consistent with those from \cite{2007ApJ...671..153Y}. However, two distinct sequences corresponding to the red and blue groups are evident. This is because the abundance matching method tends to overestimate the halo masses of blue groups and underestimate those of red groups. The deviation observed at the high-mass end is primarily due to the different HMFs used in the abundance matching method by \cite{2007ApJ...671..153Y} and in our work.

\section{DISCUSSION} \label{sec:discussion}

In this study, we use the L-GALAXIES semi-analytic model as the training and testing dataset for our ML model to predict halo masses. While employing multiple simulations for cross-validation is typically preferred for assessing the robustness of ML models, there are several reasons for our decision to focus on L-GALAXIES.

One key motivation for using L-GALAXIES is its ability to reproduce the SHMRs of red and blue galaxy groups in agreement with weak lensing observations. For instance, \cite{2024arXiv240807743W} compared the SHMRs derived from various simulations, including IllustrisTNG, Illustris, EAGLE, and L-GALAXIES, with those obtained from weak lensing. As shown in Figure 1 of \cite{2024arXiv240807743W}, L-GALAXIES provides SHMRs that most closely match the results from weak lensing, indicating that it is a suitable simulation for training ML models aimed at predicting halo masses. Given that our results may depend on the choice of simulation, we plan to incorporate additional simulations for further testing in future work.



The fact that halo masses derived from our ML model trained on L-GALAXIES not only produce SHMRs that closely align with weak lensing observations but also yield an HMF consistent with theoretical predictions suggests that, statistically, the halo masses estimated by the ML model trained on L-GALAXIES are relatively accurate.

\section{SUMMARY} \label{sec:summary}
In the $\Lambda$CDM cosmology, galaxies form and evolve in their host DM halos. Understanding the relationship between galaxies and their halos depends on accurate halo mass estimations. While the AM method has been widely used for estimating halo mass, it has been noted for its tendency to underestimate the halo masses of red groups and overestimate those of blue groups. As a consequence, the AM method derived halo masses cannot reproduce the weak lensing measurements for the red and blue groups. This paper introduces a technique for precisely estimating halo mass based on the observable properties of galaxy groups using ML method.

The setup and techniques of the ML method are summarized as follows:
\begin{itemize}
  \item Data. A mock catalog from L-GALAXIES are selected for the training, validation, and test sets, along with the \citealp{2007ApJ...671..153Y} SDSS group catalog for halo mass estimation. Key group and galaxy properties available in both the SDSS data and the mock catalog are chosen as the features for the ML model. See Section \ref{sec:data}.
  \item Sample selection effect. Since the target SDSS spectroscopic sample is flux-limited with a strong selection in stellar mass within the sample redshift range, we process the mock catalog to mimic the selection effect in observations. We implement redshift-dependent stellar mass thresholds to divide both the mock catalog and the SDSS data into complete and incomplete subsamples (see Figure \ref{fig:samples}). For the complete subsample, we use the properties of galaxies more massive than the threshold to calculate the group properties. For the incomplete subsample, the mock catalog is resampled to align the $M_{\mathrm{cen}}$ distribution with the SDSS data and we use the properties of the central galaxies instead of the properties of the groups. See Section \ref{subsubsec:sample_division}.

  \item Adding noise and avoiding overfitting. To better mimic SDSS observations and address the overfitting problem, we estimate the observation uncertainties for each observed quantity in the SDSS data, and add the same level of noise to the corresponding quantities in the mock catalog. For those quiescent galaxies in the mock catalog with SFR around or below the observation limit, we assign them new stochastic SFRs around the observation limit. See Section \ref{subsubsec:adding_noise}, Section \ref{sec:noise_importance} and Figure \ref{fig:noise_com}.
  \item Data reprocessing and normalization. Because the red group and blue group have different SHMRs, and in order to facilitate the normalization process, we have divided the groups into the red groups and the blue groups. To ensure consistency between the mock catalog and SDSS data, each quantity is normalized to follow a similar probability distribution function, as the distributions of values often do not match well between the two datasets. This normalization is performed for all observable quantities and separately for red and blue groups. As a result, the similarity in the distribution of group properties between the mock catalog and SDSS data ensures that the application of the trained model to the SDSS data does not introduce significant biases. See Section \ref{subsubsec:normalization}, Figure \ref{fig:com_distribution_blue}, Figure \ref{fig:com_distribution_red} and Figure \ref{fig:incom_distribution}.
  
  \item Machine Learning. An \texttt{XGBoost} model has been trained using the calibrated mock catalog and subsequently used to estimate halo masses in the SDSS data. Since star formation quenching of the central galaxy can decouple the coevolution of its stellar mass and halo mass, introducing substantial uncertainties in the halo mass estimation, the training has been conducted separately for blue and red groups to improve the accuracy. Due to different sample selections, the training was also conducted separately for each of the complete and incomplete subsamples at different redshift. See Section \ref{subsec:ML}.
\end{itemize}

We have tested the ML method both in the mock catalog and SDSS data. The main results are as follows:
\begin{itemize}
  \item Tests on the mock data indicate that the noise level assumed in the training set significantly impacts the accuracy of halo mass estimation. When the noise added to the training set matches the noise in the test set, the ML model produces the most precise halo mass estimates (see Figure \ref{fig:noise_com}). Therefore, as mentioned above, the best ML model is trained with a training set that has the same noise level as the measurement uncertainty in SDSS data, ensuring the most accurate halo mass estimation for SDSS groups.
  \item Tests on the mock data indicate that the AM method tends to overestimate the halo masses of blue groups and underestimate those of red groups, leading to a significant systematic bias in their relative halo masses. This effect is expected to be most pronounced at a halo mass around $\log(M_{\mathrm{halo}}/\mathrm{M_\odot})\sim12-12.5$, where there are comparable numbers of red and blue groups.  While our ML method largely eliminates the systematic bias in the derived halo masses for blue and red groups and is, on average, $\sim 1/3$ more accurate than the AM method. See Figure \ref{fig:AM}, Figure \ref{fig:incomplete}, Table \ref{table:complete_error} and Table \ref{table:incomplete_error}.
  \item With proper treatment of the compete and incomplete samples, and separated training processes, our ML method can estimate halo mass relatively accurately in the low halo mass region, down to $\log(M_{\mathrm{halo}}/\mathrm{M_\odot})\sim11.5$ or even lower, see Figure \ref{fig:incomplete} and Table \ref{table:incomplete_error}.
  \item The ML method derived HMF in SDSS closely matches those in the mock catalog and theoretical prediction in the literature (see Figure \ref{fig:HMF}). Unlike that the HMF derived form the AM method, which agrees well with the theoretical one by construction, our ML method estimates the halo mass for each individual SDSS group independently without using theoretical HMF as an input or prior. 

  \item The ML method derived SHMRs for both red and blue groups matches well with those obtained from direct weak lensing measurements. Both measurements consistently show that at a given stellar mass of the central galaxies, the halo masses of red groups are systematically more massive than those of blue groups. Conversely, the AM method derived halo masses cannot reproduce the weak lensing measurements and show little difference in the SHMR between red and blue groups. See Figure.\ref{fig:weak_lensing}.

  \item Using the halo masses for the SDSS groups derived from our ML method, the derived distribution function (i.e., HMF) agree well with theoretical predictions, and the scaling relations (i.e., SHMRs) for red and blue groups match excellently with direct weak lensing measurements, strongly supporting the that (1) the \cite{2007ApJ...671..153Y} SDSS groups are accurate; (2) sample incompleteness has been properly corrected; and (3) our ML method provides accurate halo mass estimations for individual SDSS groups.

\end{itemize}

These more accurate halo mass estimates enable a more accurate investigation of the galaxy-halo connection and other important halo-related science, such as galactic conformity and the role of halos in galaxy formation and evolution. This method can be applied to forthcoming large surveys to higher redshifts, such as MOONS \citep{2014SPIE.9147E..0NC} and CSST \citep{2011SSPMA..41.1441Z}.

\acknowledgments
Y.P. and D.Z. acknowledge National Science Foundation of China (NSFC) Grant No. 12125301, 12192220, 12192222, and the science research grants from the China Manned Space Project with No. CMS-CSST-2021-A07. LCH was supported by the National Science Foundation of China (11991052, 12233001), the National Key R\&D Program of China (2022YFF0503401), and the China Manned Space Project (CMS-CSST-2021-A04, CMS-CSST-2021-A06). QSGU is supported by the National Natural Science Foundation of China (No. 12192222, 12192220 and 12121003).

\section*{Data Availability}
All derived data in this work will be fully publicly available in our following data release paper, together with various other derived properties of the halos and galaxies.

\clearpage

\appendix

\begin{deluxetable}{cccccccccccccc}[htbp]
  \tablecaption{Estimation errors of complete sample in test set\label{table:complete_error}}
  \tablehead{
  \colhead{$i$} & \colhead{$\sigma_{n}/\sigma_0$} & 
  \colhead{$\mu_{AM}^{b}$} & \colhead{$\sigma_{AM}^{b}$} & 
  \colhead{$\mu_{AM}^{r}$} & \colhead{$\sigma_{AM}^{r}$} & 
  \colhead{$\mu_{a}^{b}$} & \colhead{$\sigma_{a}^{b}$} & 
  \colhead{$\mu_{a}^{r}$} & \colhead{$\sigma_{a}^{r}$} & \colhead{$\mu_{na}^{b}$} & \colhead{$\sigma_{na}^{b}$} & \colhead{$\mu_{na}^{r}$} &\colhead{$\sigma_{na}^{r}$} \\ 
  \colhead{(1)} & \colhead{(2)} & \colhead{(3)} & \colhead{(4)} & 
  \colhead{(5)} & \colhead{(6)} & \colhead{(7)} &
  \colhead{(8)} & \colhead{(9)} & \colhead{(10)} & \colhead{(11)} &\colhead{(12)} & \colhead{(13)} &\colhead{(14)}
  } 
  \startdata
  0     & 0     & 0.045  & 0.155  & -0.127  & 0.245  & 0.000  & 0.082  & 0.000  & 0.153  & 0.000  & 0.087  & -0.001  & 0.166  \\
    & 0.5   & 0.045  & 0.160  & -0.128  & 0.250  & 0.000  & 0.098  & 0.001  & 0.195  & 0.000  & 0.099  & 0.000  & 0.217  \\
    & 1     & 0.044  & 0.171  & -0.129  & 0.260  & 0.000  & 0.116  & 0.002  & 0.217  & 0.000  & 0.117  & -0.002  & 0.225  \\
    & 2     & 0.041  & 0.206  & -0.104  & 0.297  & 0.000  & 0.152  & 0.001  & 0.236  & -0.001  & 0.152  & -0.002  & 0.240  \\\hline
  1     & 0     & 0.107  & 0.201  & -0.137  & 0.270  & 0.000  & 0.108  & 0.001  & 0.158  & 0.000  & 0.114  & 0.000  & 0.175  \\
    & 0.5   & 0.106  & 0.209  & -0.139  & 0.272  & 0.000  & 0.124  & 0.000  & 0.199  & 0.000  & 0.127  & -0.001  & 0.230  \\
    & 1     & 0.103  & 0.222  & -0.139  & 0.286  & 0.000  & 0.146  & 0.001  & 0.226  & 0.000  & 0.150  & -0.002  & 0.239  \\
    & 2     & 0.093  & 0.267  & -0.115  & 0.334  & -0.001  & 0.188  & -0.001  & 0.257  & 0.000  & 0.190  & 0.000  & 0.261  \\\hline
  2     & 0     & 0.160  & 0.243  & -0.130  & 0.304  & 0.000  & 0.133  & 0.000  & 0.177  & 0.001  & 0.135  & -0.001  & 0.195  \\
    & 0.5   & 0.160  & 0.252  & -0.132  & 0.309  & 0.000  & 0.151  & -0.001  & 0.219  & 0.001  & 0.154  & 0.000  & 0.248  \\
    & 1     & 0.152  & 0.268  & -0.129  & 0.322  & 0.000  & 0.172  & 0.002  & 0.245  & 0.000  & 0.181  & 0.001  & 0.260  \\
    & 2     & 0.132  & 0.319  & -0.103  & 0.377  & 0.005  & 0.224  & -0.002  & 0.278  & 0.002  & 0.226  & 0.001  & 0.285  \\\hline
  3     & 0     & 0.211  & 0.289  & -0.124  & 0.337  & -0.003  & 0.151  & -0.002  & 0.200  & 0.000  & 0.157  & -0.001  & 0.220  \\
    & 0.5   & 0.209  & 0.289  & -0.123  & 0.342  & 0.003  & 0.173  & -0.003  & 0.239  & 0.002  & 0.177  & -0.001  & 0.268  \\
    & 1     & 0.198  & 0.310  & -0.113  & 0.357  & 0.002  & 0.202  & 0.001  & 0.267  & 0.001  & 0.212  & 0.001  & 0.281  \\
    & 2     & 0.153  & 0.359  & -0.079  & 0.419  & -0.005  & 0.260  & -0.003  & 0.303  & 0.003  & 0.259  & -0.002  & 0.308  \\\hline
  4     & 0     & 0.294  & 0.373  & -0.123  & 0.427  & -0.004  & 0.211  & 0.005  & 0.247  & 0.005  & 0.212  & -0.006  & 0.272  \\
    & 0.5   & 0.304  & 0.370  & -0.121  & 0.442  & 0.003  & 0.222  & -0.007  & 0.301  & 0.006  & 0.237  & 0.002  & 0.320  \\
    & 1     & 0.252  & 0.417  & -0.088  & 0.467  & -0.013  & 0.280  & -0.001  & 0.327  & 0.002  & 0.267  & -0.007  & 0.350  \\
    & 2     & 0.195  & 0.443  & -0.060  & 0.510  & 0.007  & 0.323  & 0.003  & 0.370  & 0.010  & 0.328  & 0.002  & 0.377  \\\hline
  5     & 0     & 0.355  & 0.419  & -0.084  & 0.483  & -0.009  & 0.264  & 0.005  & 0.282  & -0.018  & 0.283  & 0.000  & 0.312  \\
    & 0.5   & 0.333  & 0.443  & -0.064  & 0.490  & -0.004  & 0.289  & 0.006  & 0.329  & -0.013  & 0.295  & 0.000  & 0.349  \\
    & 1     & 0.344  & 0.442  & -0.060  & 0.537  & 0.023  & 0.321  & -0.004  & 0.363  & 0.026  & 0.339  & -0.013  & 0.391  \\
    & 2     & 0.218  & 0.552  & -0.040  & 0.576  & 0.004  & 0.424  & -0.001  & 0.426  & 0.010  & 0.425  & -0.009  & 0.450  \\
  \enddata
  \tablecomments{(1) The sample number, see Table \ref{tab:error}. (2) The amount of noise added to the training and test set with the measurement errors in observation as the unit. (3) The mean of residuals in AM method for blue groups. (4) The standard deviation of residuals in AM method for blue groups. (5) The mean of residuals in AM method for red groups. (6) The standard deviation of residuals in AM method for red groups. (7) The mean of residuals in ML method with age as a feature for blue groups. (8) The standard deviation of residuals in ML method with age as a feature for blue groups. (9) The mean of residuals in ML method with age as a feature for red groups. (10) The standard deviation of residuals in ML method with age as a feature for red groups. (11) The mean of residuals in ML method without age as a feature for blue groups. (12) The standard deviation of residuals in ML method without age as a feature for blue groups. (13) The mean of residuals in ML method without age as a feature for red groups. (14) The standard deviation of residuals in ML method without age as a feature for red groups.}
  
  \end{deluxetable}

  \begin{deluxetable}{cccccccccc}[htbp]
  \tablecaption{Estimation error in incomplete sample\label{table:incomplete_error}}
  \tablehead{
  \colhead{$i$} & \colhead{$\sigma_{n}/\sigma_0$} & 
  \colhead{$\mu_{a}^{b}$} & \colhead{$\sigma_{a}^{b}$} & 
  \colhead{$\mu_{a}^{r}$} & \colhead{$\sigma_{a}^{r}$} & \colhead{$\mu_{na}^{b}$} & \colhead{$\sigma_{na}^{b}$} & \colhead{$\mu_{na}^{r}$} &\colhead{$\sigma_{na}^{r}$} \\ 
  \colhead{(1)} & \colhead{(2)} & \colhead{(3)} & \colhead{(4)} & 
  \colhead{(5)} & \colhead{(6)} & \colhead{(7)} &
  \colhead{(8)} & \colhead{(9)} & \colhead{(10)}
  } 
  \startdata
  1     & 0     & 0.000  & 0.080  & -0.002  & 0.189  & 0.000  & 0.084  & 0.000  & 0.198  \\
        & 0.5   & 0.000  & 0.094  & 0.002  & 0.226  & 0.001  & 0.096  & 0.003  & 0.257  \\
        & 1     & 0.000  & 0.113  & -0.004  & 0.254  & 0.000  & 0.114  & -0.002  & 0.267  \\
        & 2     & -0.001  & 0.150  & -0.002  & 0.261  & 0.000  & 0.151  & 0.003  & 0.261  \\\hline
  2     & 0     & 0.000  & 0.100  & 0.000  & 0.188  & 0.000  & 0.102  & -0.003  & 0.202  \\
        & 0.5   & 0.000  & 0.113  & 0.002  & 0.227  & 0.000  & 0.116  & 0.001  & 0.250  \\
        & 1     & 0.000  & 0.132  & 0.000  & 0.254  & 0.000  & 0.135  & 0.004  & 0.262  \\
        & 2     & -0.001  & 0.175  & -0.001  & 0.272  & 0.001  & 0.176  & 0.004  & 0.281  \\\hline
  3     & 0     & 0.000  & 0.117  & 0.001  & 0.214  & -0.001  & 0.123  & 0.001  & 0.222  \\
        & 0.5   & 0.002  & 0.138  & 0.005  & 0.247  & 0.000  & 0.139  & 0.004  & 0.276  \\
        & 1     & 0.001  & 0.160  & 0.002  & 0.270  & -0.001  & 0.162  & -0.002  & 0.282  \\
        & 2     & 0.003  & 0.207  & -0.007  & 0.301  & -0.003  & 0.212  & -0.006  & 0.311  \\\hline
  4     & 0     & -0.002  & 0.169  & 0.009  & 0.252  & 0.002  & 0.164  & 0.006  & 0.279  \\
        & 0.5   & 0.000  & 0.189  & 0.008  & 0.300  & 0.004  & 0.187  & -0.002  & 0.322  \\
        & 1     & 0.000  & 0.214  & -0.005  & 0.320  & 0.004  & 0.213  & 0.002  & 0.335  \\
        & 2     & 0.003  & 0.266  & 0.007  & 0.357  & 0.000  & 0.270  & 0.008  & 0.367  \\\hline
  5     & 0     & 0.007  & 0.183  & -0.010  & 0.308  & -0.007  & 0.202  & -0.008  & 0.331  \\
        & 0.5   & -0.002  & 0.233  & -0.009  & 0.347  & 0.009  & 0.229  & 0.003  & 0.375  \\
        & 1     & -0.004  & 0.269  & -0.001  & 0.376  & -0.012  & 0.271  & -0.015  & 0.395  \\
        & 2     & 0.013  & 0.309  & 0.006  & 0.407  & -0.008  & 0.336  & -0.020  & 0.440  \\
  \enddata
  \tablecomments{(1) The sample number, see Table \ref{tab:error}. (2) The amount of noise added to the training and test set with the measurement errors in observation as the unit. (3) The mean of residuals in ML method with age as a feature for blue groups. (4) The standard deviation of residuals in ML method with age as a feature for blue groups. (5) The mean of residuals in ML method with age as a feature for red groups. (6) The standard deviation of residuals in ML method with age as a feature for red groups. (7) The mean of residuals in ML method without age as a feature for blue groups. (8) The standard deviation of residuals in ML method without age as a feature for blue groups. (9) The mean of residuals in ML method without age as a feature for red groups. (10) The standard deviation of residuals in ML method without age as a feature for red groups.}
  \end{deluxetable}

\section{Estimation error in test set}
We present in Table \ref{table:complete_error} the prediction errors of the ML method and AM method on the complete sample of the test set across different redshift bins. Table \ref{table:incomplete_error} shows the prediction errors of the ML method on the incomplete sample of the test set across different redshift bins.

\clearpage

\begin{figure*}[t]
  \centering
  \includegraphics[width=0.9\textwidth]{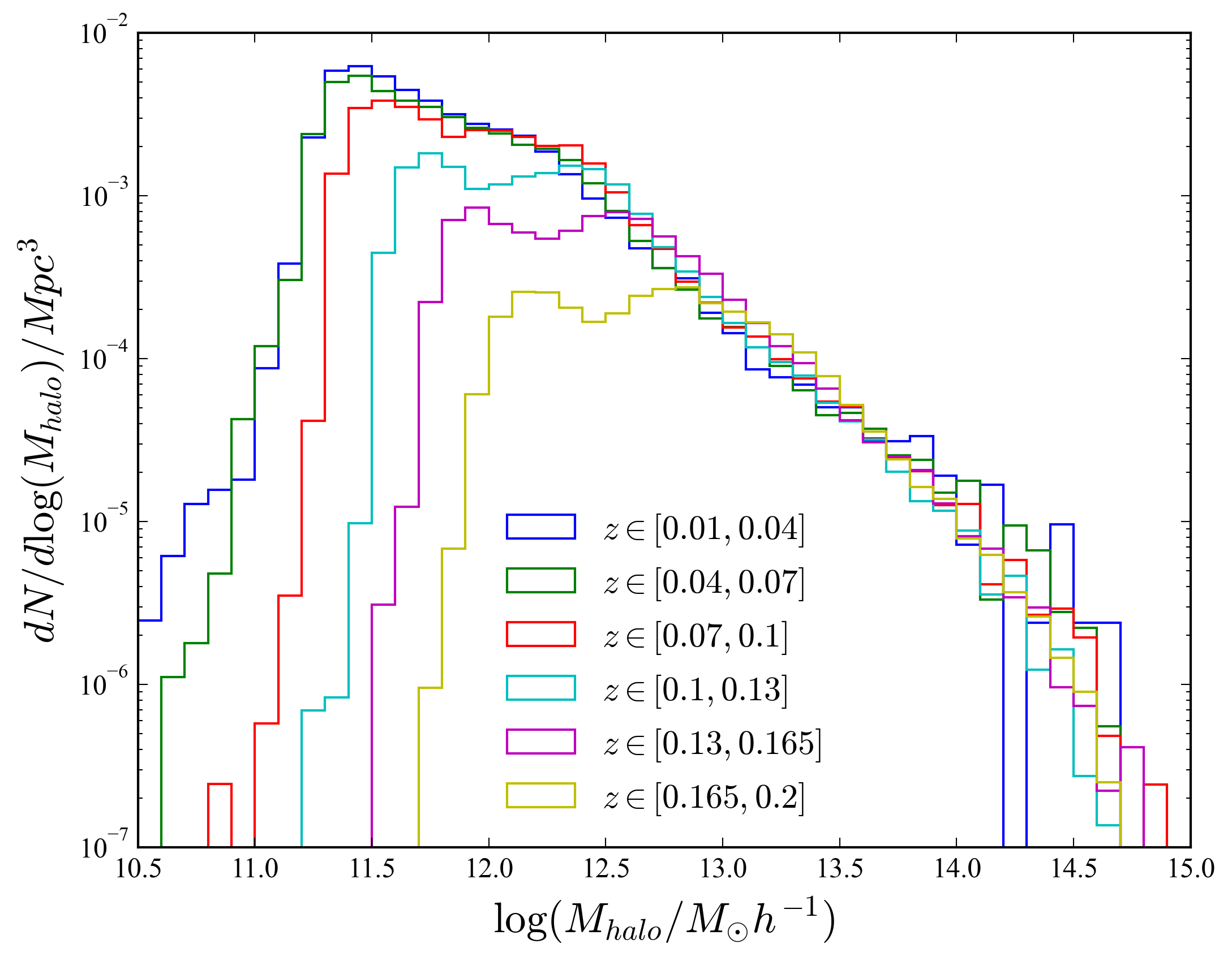}
  \caption{HMFs in different redshift bins. HMFs derived from SDSS data using ML models trained on training sets with Gaussian noise levels of $\sigma_n/\sigma_0=1$. All HMFs have been corrected using $V_{max}$ correction.}
  \label{fig:hmf_redshift}
\end{figure*}

\section{HMFs in different redshift bins}\label{sec:hmf_redshift}

In Figure \ref{fig:hmf_redshift}, we present the HMFs for the six redshift bins. It can be seen that at the high-mass end, the HMFs across different redshift bins are remarkably consistent. This consistency indicates that our estimated halo masses do not exhibit significant redshift-dependent systematic biases.

Figure \ref{fig:hmf_redshift} shows the HMFs in different redshift bins. For the two redshift intervals [0.01,0.04] and [0.04,0.07], the HMFs are almost identical. This indicates that, after applying the $V_{max}$ correction, the group catalogs in these two bins can be considered complete. The decline at the low-mass end of the HMF is due to our selection criterion: we only include groups whose central galaxy stellar masses are greater than $10^{9.5}M_\odot$. From the SHMRs shown in Figure 9 of our paper, we find that when the central galaxy mass is $10^{9.5}M_\odot$, the corresponding halo mass is approximately $10^{11.5}M_\odot$. Therefore, selecting groups with central galaxy stellar masses above $10^{9.5}M_\odot$ does not affect the HMF for halos with masses greater than $10^{11.5}M_\odot$. Consequently, in these two redshift bins, we can consider the HMF to be complete and reliable for $M_\mathrm{halo}>10^{11.5}M_\odot$.

For higher redshift bins, since the group catalog is flux-limited, the HMF starts to be affected by the incompleteness of the group catalog. In these cases, the completeness limit of the HMF corresponds to the halo mass where the HMF begins to deviate from those of the lower redshift bins. 

\clearpage
\bibliography{sample63}{}
\bibliographystyle{aasjournal}



\end{document}